\documentclass{article}
\usepackage{bookstyle,graphicx,bm,amsmath,amssymb}

\newcommand{\Eq}[1]{Eq.~\eqref{#1}}
\newcommand{\pdag}{{\phantom{\dagger}}}
\newcommand{\pprime}{{\phantom{\prime}}}
\newcommand{\past}{{\phantom{\ast}}}

\newcommand{\beq}{\begin{equation}}
\newcommand{\eeq}{\end{equation}}

\newcommand{\beqa}{\begin{eqnarray}}
\newcommand{\eeqa}{\end{eqnarray}}

\newcommand{\ket}[1]{\left|{#1}\right\rangle}

\newcommand{\PRB}{Phys. Rev. B~} 
\newcommand{\PRL}{Phys. Rev. Lett.~} 
\newcommand{\RMP}{Rev. Mod. Phys.~} 

\DeclareMathOperator{\sign}{sign}
\DeclareMathOperator{\im}{Im}

\begin{document}

\author{Leonid I. Glazman$^1$ and Michael Pustilnik$^2$}
\runauthor{L.I. Glazman and M. Pustilnik}
\address{$^1$William I. Fine Theoretical Physics Institute, 
University of Minnesota, \\
Minneapolis, MN 55455, USA \\
$^2$School of Physics, Georgia Institute of Technology, \\
Atlanta, GA 30332, USA}
\title{Low-temperature transport through a quantum dot}
\runtitle{Low-temperature transport through a quantum dot}
\maketitle

\begin{abstract}
We review mechanisms of low-temperature electronic transport 
through a quantum dot weakly coupled to two conducting leads. 
Transport in this case is dominated by electron-electron 
interaction. At temperatures moderately lower than the charging 
energy of the dot, the linear conductance is suppressed by the 
Coulomb blockade. Upon further lowering of the temperature, 
however, the conductance may start to increase again due to the 
Kondo effect. We concentrate on lateral quantum dot systems 
and discuss the conductance in a broad temperature range, which 
includes the Kondo regime. 
\end{abstract}

\clearpage
\section{Introduction}

In quantum dot devices~\cite{blockade} a small droplet of electron
liquid, or just a few electrons are confined to a finite region of
space. The dot can be attached by tunneling junctions to massive
electrodes to allow electronic transport across the system.  The
conductance of such a device is determined by the number of 
electrons on the dot $N$, which in turn is controlled by varying 
the potential on the gate - an auxiliary electrode capacitively coupled 
to the dot~\cite{blockade}. At sufficiently low temperatures, $N$ 
is an integer at almost any gate voltage $V_g$. Exceptions are 
narrow intervals of $V_g$ in which an addition of a single electron 
to the dot does not change much the electrostatic energy of the system. 
Such a degeneracy between different charge states of the dot allows 
for an activationless electron transfer through it, whereas for all other 
values of $V_g$ the activation energy for the conductance $G$ 
across the dot is finite. The resulting oscillatory dependence 
$G(V_g)$ is the hallmark of the Coulomb blockade 
phenomenon~\cite{blockade}. The contrast between the low- 
and high-conductance regions (Coulomb blockade valleys and 
peaks, respectively) gets sharper at lower temperatures. The pattern 
of periodic oscillations in the $G$ vs. $V_g$ dependence is 
observed down to the lowest attainable temperatures in experiments 
on tunneling through small metallic islands~\cite{Devoret}. 

Conductance through quantum dots formed in semiconductor 
heterostructures exhibits a reacher behavior~\cite{blockade}. In 
larger dots (in the case of GaAs heterostructures, such dots may 
contain hundreds of electrons), the fluctuations of the heights of 
the Coulomb blockade peaks become apparent already at moderately 
low temperatures. Characteristic for mesoscopic phenomena, the 
heights are sensitive to the shape of a dot and to the magnetic flux threading 
it. The separation in gate voltage between the Coulomb blockade 
peaks, and the conductance in the valleys also exhibit mesoscopic 
fluctuations. However, the pattern of sharp conductance
peaks separating the low-conductance valleys of the $G(V_g)$
dependence persists. Smaller quantum dots (containing few tens of 
electrons in the case of GaAs) show yet another feature~\cite{kondo_exp}: 
in some Coulomb blockade valleys the dependence $G(T)$ is not 
monotonic and has a minimum at a finite temperature. This minimum 
is similar in origin~\cite{kondo_popular} to the well-known non-monotonic
temperature dependence of the resistivity of a metal containing
magnetic impurities~\cite{Kondo} -- the \textit{Kondo effect}. 
Typically, the valleys with anomalous temperature dependence 
correspond to an odd number of electrons in the dot. In an
ideal case, the low-temperature conductance in such a valley 
is of the order of conductance at peaks surrounding it. Thus, 
at low temperatures the two adjacent peaks merge to form a 
broad maximum.

The number of electrons on the dot is a well-defined quantity 
as long the conductances of the junctions connecting the dot to the
electrodes is small compared to the conductance quantum $e^2/h$. 
In quantum dot devices formed in semiconductor heterostructures 
the conductances of junctions can be tuned continuously. With 
the increase of the conductances, the periodic pattern in $G(V_g)$ 
dependence gradually gives way to mesoscopic conductance 
fluctuations. Yet, electron-electron interaction still affects the 
transport through the device.  A strongly asymmetric quantum 
dot device with one junction weakly conducting, while another 
completely open, provides a good example of that~\cite{FM,AG1,Cronenwett}. 
The differential conductance across the device in this case exhibits 
zero-bias anomaly -- suppression at low bias. Clearly, Coulomb
blockade is not an isolated phenomenon, but is closely related to 
interaction-induced anomalies of electronic transport and 
thermodynamics in higher dimensions~\cite{Altsuler_Aronov}.

The emphasis of these lectures is on the Kondo effect in quantum 
dots. We will concentrate on the so-called \textit{lateral quantum 
dot systems}~\cite{blockade,kondo_exp}, formed by gate 
depletion of a two-dimensional electron gas at the interface between 
two semiconductors. These devices offer the highest degree of 
tunability, yet allow for relatively simple theoretical treatment. At 
the same time, many of the results presented below are directly 
applicable to other systems as well, including  vertical quantum 
dots~\cite{vertical,Sasaki,induced_review}, Coulomb-blockaded 
carbon nanotubes~\cite{induced_review,nanotube}, single-molecule 
transistors~\cite{Park}, and magnetic atoms on metallic 
surfaces~\cite{mirage}. 

Kondo effect emerges at relatively low temperature, and we will follow
the evolution of the conductance upon the reduction of temperature. On
the way to Kondo effect, we encounter also the phenomena of Coulomb
blockade and of mesoscopic conductance fluctuations.

\section{Model of a lateral quantum dot system} 
\label{model}

The Hamiltonian of interacting electrons confined to a quantum dot 
has the following general form,
\beq
H_{\rm dot}= 
\sum_s \sum_{ij}
h^\pdag_{ij}d^\dagger_{is}d^\pdag_{js}
+
\frac{1}{2}\sum_{s s'}\sum_{ijkl} h^\pdag_{ijkl}
d^\dagger_{i s} d^\dagger_{j s'} 
d^\pdag_{k s'}d^\pdag_{l s}.
\label{2.1}
\eeq
Here an operator $d^\dagger_{is}$ creates an electron with spin 
$s$ in the orbital state $\phi_i(\bm r)$ (the wave functions are 
normalized according to
$\int d\bm r \phi^\ast_i (\bm r) \phi^\past_j (\bm r) = \delta_{ij})$; 
$h_{ij}^\past=h^*_{ji}$ is an Hermitian matrix describing the 
single-particle part of the Hamiltonian. The matrix elements 
$h_{ijkl}$ depend on the potential $U(\bm r-\bm r')$ of 
electron-electron interaction,
\beq
h_{ijkl}
=\int d\bm r \, d\bm r' 
\phi_i^*(\bm r) \phi_j^*(\bm r')
U(\bm r-\bm r')
\phi_k^\past (\bm r') \phi_l^\past (\bm r).
\label{2.2}
\eeq

The Hamiltonian~\eqref{2.1} can be simplified further provided 
that the quasiparticle spectrum is not degenerate near the Fermi 
level, that the Fermi-liquid theory is applicable to the description 
of the dot, and that the dot is in the metallic conduction regime. 
The first of these conditions is satisfied if the dot has no spatial 
symmetries, which implies also that motion of quasiparticles within 
the dot is chaotic. 

The second condition is met if the electron-electron interaction 
within the dot is not too strong, i.e. the gas parameter $r_s$ 
is small, 
\beq
r_s = (k_F a_0)^{-1} \lesssim 1,
\quad
a_0 = \kappa \hbar^2/e^2 m^\ast 
\label{2.3}
\eeq
Here $k_F$ is the Fermi wave vector, $a_0$ is the effective 
Bohr radius, $\kappa$ is the dielectric constant of the material, 
and $m^\ast$ is the quasiparticle effective mass.

The third condition requires the ratio of the Thouless energy 
$E_T$ to the mean single-particle level spacing $\delta E$ to 
be large~\cite{RMT1}, 
\beq
g = E_T/\delta E \gg 1.
\label{2.4}
\eeq
For a ballistic two-dimensional dot of linear size $L$ the Thouless 
energy $E_T$ is of the order of $\hbar v_F/L$, whereas the level 
spacing can be estimated as 
\beq
\delta E \sim \hbar v_F k_F/N \sim \hbar^2/m^\ast L^2 .
\label{2.5}
\eeq
Here $v_F$ is the Fermi velocity and $N\sim (k_F L)^2$ is the 
number of electrons in the dot. Therefore,
\beq
g \sim k_F L \sim \sqrt N\,,
\label{2.4a}
\eeq
so that having a large number of electrons $N\gg 1$ in the dot 
guarantees that the condition~\Eq{2.4} is satisfied.  

Under the conditions~\eqref{2.3},~\eqref{2.4} the 
\textit{Random Matrix Theory} (for a review see, 
e.g.,~\cite{Beenakker_RMP,Alhassid_RMP,Haake,Efetov}) 
is a good starting point for description of non-interacting 
quasiparticles within the energy strip of the width $E_T$ about 
the Fermi level~\cite{RMT1}. The matrix elements $h_{ij}$ in 
\Eq{2.1} belong to a Gaussian ensemble~\cite{Haake,Efetov}. Since 
the matrix elements do not depend on spin, each eigenvalue 
$\epsilon_n$ of the matrix $h_{ij}$ represents a spin-degenerate 
energy level. The spacings $\epsilon_{n+1}-\epsilon_n$ between 
consecutive levels obey the Wigner-Dyson statistics~\cite{Haake}; 
the mean level spacing 
$\langle{\epsilon_{n+1}-\epsilon_n}\rangle = \delta E$.

We discuss now the second term in the Hamiltonian~\eqref{2.1}, 
which describes electron-electron interaction. It turns 
out~\cite{RMT,KAA,ABG} that the vast majority of the matrix 
elements $h_{ijkl}$ are small. Indeed, in the lowest order in 
$1/g\ll 1$, the wave functions $\phi_i (\bm r)$ are Gaussian 
random variables with zero mean, statistically independent of each 
other and of the corresponding energy levels~\cite{wave_functions}:
\beq
\overline{\phi_i^*(\bm r)\phi_j^\past(\bm r')}
= \frac{\delta_{ij}}{\cal A}\, F(|\bm r - \bm r'|)\,,
\quad
\overline{\phi_i(\bm r)\phi_j(\bm r')} 
= \frac{\delta_{\beta,1}\delta_{ij}}{\cal A}\, F(|\bm r - \bm r'|)\,.
\label{2.6a}
\eeq
Here ${\cal A}\sim L^2$ is the area of the dot, and the function $F$ 
is given by
\beq
F(r) \sim \langle\exp(i\bm k\cdot\bm r)\rangle_{\rm FS} \,. 
\label{2.6b} 
\eeq
where $\langle\ldots\rangle_{\rm FS}$ stands for the averaging 
over the Fermi surface $|\bm k| = k_F$. In two dimensions, the 
function $F(r)$ decreases with $r$ as $F\propto (k_F r)^{-1/2}$ 
at $k_F r\gg 1$, and saturates to $F\sim 1$ at $k_F r\ll 1$. 

The parameter $\beta$ in \Eq{2.6a} distinguishes between the presence 
$(\beta = 1)$ or absence $(\beta = 2)$ of the time-reversal symmetry.
The symmetry breaking is driven by the orbital effect of the magnetic field 
and is characterised by the parameter
\[
\chi = (\Phi/\Phi_0)\sqrt{g}, 
\]
where $\Phi$ is the magnetic flux threading the dot and $\Phi_0 =
hc/e$ is the flux quantum, so that the limits $\chi\ll 1$ and $\chi\gg
1$ correspond to, respectively, $\beta=1$ and $\beta=2$. Note that in
the case of a magnetic field $H_\perp$ applied perpendicular to the
plane of the dot, the crossover (at $\chi\sim 1$) between the two
regimes occurs at so weak field that the corresponding Zeeman energy
$B$ is negligible\footnote{For example, in the
  experiments~\cite{Folk96} the crossover takes place at $H_\perp\sim
  10~mT$. Zeeman energy in such a field $B\sim 2.5~mK$, which is by an
  order of magnitude lower than the base temperature in the
  measurements.}.

After averaging with the help of Eqs.~\eqref{2.6a}-\eqref{2.6b}, the matrix
elements~\eqref{2.2} take the form
\[
\overline{h_{ijkl}}
= \bigl(2E_C + E_S/2\bigr)\delta_{il}\delta_{jk} + E_S\delta_{ik}\delta_{jl}
+ \Lambda\bigl({2}/{\beta} - 1\bigr) \delta_{ij}\delta_{kl} .
\]
We substitute this expression into Hamiltonian~\eqref{2.1}, and 
rearrange the sum over the spin indexes with the help of the identity 
\beq
2\,\delta_{s_1^\pprime s_2^\pprime} \delta_{s_1^\prime s_2^\prime} 
= \delta_{s_1^\pprime s_1^\prime} \delta_{s_2^\prime s_2^\pprime} 
+ \bm\sigma_{s_1^\pprime s_1^\prime} \cdot
\bm\sigma_{s_2^\prime s_2^\pprime} ,
\label{0}
\eeq
where $\bm\sigma = (\sigma^x,\sigma^y,\sigma^z)$ are the Pauli 
matrices. This results in a remarkably simple form~\cite{KAA,ABG}
\beq
H_{\rm int} = E_C\hat{N}^2 - E_S \hat{\bf S}^2
+\Lambda\bigl({2}/{\beta} - 1\bigr)\hat{T}^\dagger\hat{T} 
\label{2.7}
\eeq
of the interaction part of the Hamiltonian of the dot. Here 
\beq
 \hat{N} =\sum_{ns} d^\dagger_{ns}d^\pdag_{ns},
\quad
\hat{\bf S} = \sum_{nss'}
d^\dagger_{ns}\frac{{\bm\sigma}_{ss'}}{2}\,d^\pdag_{ns'}
\quad
\hat{T}=\sum_n d^\dagger_{n\uparrow}d^\dagger_{n\downarrow}
\label{2.8}
\eeq
are the operators of the total number of electrons in the dot, 
of the dot's spin, and the ``pair creation'' operator
corresponding to the interaction in the Cooper channel.\\

\begin{figure}[h]
\centerline{\includegraphics[width=0.45\textwidth]{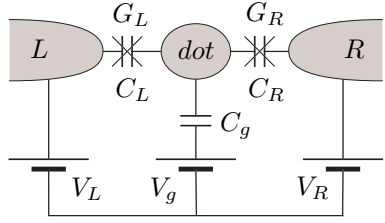}}
\vspace{2.5 mm}
\caption{Equivalent circuit for a quantum dot connected to two 
leads by tunnel junctions and capacitively coupled to the gate electrode. 
The total capacitance of the 
dot $C=C_L+C_R+C_g$.}
\label{circuit}
\end{figure}

The first term in \Eq{2.7} represents the electrostatic energy. 
In the conventional equivalent circuit picture, see Fig.~\ref{circuit}, 
the charging energy $E_C$ is related to the total capacitance $C$ 
of the dot, $E_C=e^2/2C$. For a mesoscopic ($k_F L\gg 1$) 
conductor, the charging energy is large compared to the mean level 
spacing $\delta E$. Indeed, using the estimates $C\sim\kappa L$ 
and~Eqs.~\eqref{2.3} and \eqref{2.5}, we find
\beq
E_C/\delta E \sim L/a_0 
\sim r_s \sqrt N .
\label{2.9}
\eeq
Except an exotic case of an extremely weak interaction, this ratio is 
large for $N\gg 1$; for the smallest quantum dots formed in GaAs 
heterostructures, $E_C /\delta E\sim 10$~\cite{kondo_exp}. Note that
Eqs.~\eqref{2.4}, \eqref{2.4a}, and \eqref{2.9} imply that
\[
E_T/E_C \sim 1/r_s \gtrsim 1,
\]
which justifies the use of RMT for the description of single-particle 
states with energies $|\epsilon_n|\lesssim E_C$, relevant for 
Coulomb blockade.
 
The second term in \Eq{2.7} describes the intra-dot exchange 
interaction, with the exchange energy $E_S$ given by
\beq
E_S = \int d\bm r\, d\bm r' U(\bm r-\bm r') F^2(|\bm r - \bm r'|) 
\label{2.10}
\eeq
In the case of a long-range interaction the potential $U$ here 
should properly account for the screening~\cite{ABG}. For 
$r_s\ll 1$ the exchange energy can be estimated with logarithmic 
accuracy by substituting $U(r) = (e^2/\kappa r)\theta(a_0 - r)$ 
into \Eq{2.10} (here we took into account that the screening 
length in two dimensions coincides with the Bohr radius $a_0$), 
which yields
\beq
E_S \sim r_s\ln\left(1/r_S\right)\delta E \ll \delta E. 
\label{2.11}
\eeq
The estimate~\Eq{2.11} is valid only for $r_s\ll 1$. However, 
the ratio $E_S/\delta E$ remains small for experimentally
relevant\footnote{For GaAs ($m^*\approx 0.07 m_e$, 
$\kappa\approx 13$) the effective Bohr radius $a_0\approx 10~nm$, 
whereas a typical density of the two-dimensional electron gas,
$n\sim10^{11}~cm^{-2}$~\cite{kondo_exp}, corresponds to 
$k_F = \sqrt{2\pi n}\sim 10^6~cm^{-1}$. This gives 
$k_F a_0 \sim 1$.}
value $r_s\sim 1$ as long as the Stoner criterion for the 
absence of itinerant magnetism~\cite{Ziman} is satisfied. 
This guarantees the absence of a macroscopic (proportional 
to $N$) magnetization of a dot in the ground state~\cite{KAA}.

The third term in \Eq{2.7}, representing interaction in the Cooper 
channel, is renormalized by higher-order corrections arising due to
virtual transitions to states outside the energy strip of the width $E_T$
about the Fermi level. For attractive interaction ($\Lambda<0$) 
the renormalization enhances the interaction, eventually leading to 
the superconducting instability and formation of a gap $\Delta_\Lambda$ 
in the electronic spectrum. Properties of very small 
$(\Delta_\Lambda\sim\delta E)$ superconducting grains are reviewed 
in, e.g.,~\cite{Delft-Ralph}; for properties of larger grains 
$(\Delta_\Lambda\sim E_C)$ see~\cite{MGS}. 
Here we concentrate on the repulsive interaction $(\Lambda>0)$, 
in which case $\Lambda$ is very small,
\[
\Lambda\sim\frac{\delta E}{\ln (\epsilon_F/E_T)} 
\sim \frac{\delta E}{\ln N}
\ll \delta E .
\]
This estimate accounts for the logarithmic renormalization 
of $\Lambda$ when the high-energy cutoff is reduced from 
the Fermi energy $\epsilon_F$ down to the Thouless energy 
$E_T$~\cite{ABG}. In addition, if the time-reversal symmetry 
is lifted $(\beta =2)$ then the third term in \Eq{2.7} is zero to start 
with. Accordingly, hereinafter we neglect this term altogether by 
setting $\Lambda=0$. 

Obviously, the interaction part of the Hamiltonian, \Eq{2.7}, 
is invariant with respect to a change of the basis of single-particle 
states $\phi_i(\bm r)$. Picking up the basis in which the first term 
in \Eq{2.1} is diagonal, we arrive at the \textit{universal 
Hamiltonian}~\cite{KAA,ABG},
\beq
H_{\rm dot} = 
\sum_{ns}\epsilon^\pdag_n d_{ns}^\dagger d^\pdag_{ns}
+ E_C \bigl({\hat N}- N_0\bigr)^2 -E_S \hat{\bf S}^2.
\label{2.12}
\eeq
We included in \Eq{2.12} the effect of the capacitive coupling to the
gate electrode: the dimensionless parameter $N_0$ is proportional to
the gate voltage, 
\[
N_0=C_gV_g/e,
\] 
where $C_g$ is the capacitance between the dot and the gate, see
Fig.~\ref{circuit}.
The relative magnitude of fully off-diagonal interaction terms in \Eq{2.1}
(corresponding to $h_{ijkl}$ with all four indices different), not included 
in~\Eq{2.12}, is of the order of $1/g\sim N^{-1/2}\ll 1$. Partially diagonal 
terms (two out of four indices coincide) are larger, of the order of 
$\sqrt{1/g}\sim N^{-1/4}$, but still are assumed to
be negligible as $N\gg 1$. 

As discussed above, in this limit the energy scales involved 
in~\Eq{2.12} form a well-defined hierarchy 
\beq
E_S\ll\delta E\ll E_C .
\label{2.13}
\eeq
If all the single-particle energy levels $\epsilon_n$ were equidistant, 
then the spin $S$ of an even-$N$ state would be zero, while an 
odd-$N$ state would have $S=1/2$. However, the level spacings 
are random. If the spacing between the highest occupied level and 
the lowest unoccupied one is accidentally small, than the gain in the 
exchange energy, associated with the formation of a higher-spin state, 
may be sufficient to overcome the loss of the kinetic energy (cf. the 
Hund's rule in quantum mechanics). For $E_S\ll \delta E$ such 
deviations from the simple even-odd periodicity are 
rare~\cite{KAA,spin,spin_exp}. This is why the last term 
in~\Eq{2.12} is often neglected. \Eq{2.12} then reduces to the 
Hamiltonian of the \textit{Constant Interaction Model}, widely used 
in the analysis of experimental data~\cite{blockade}. Finally, it should 
be emphasized that SU(2)--invariant Hamiltonian \eqref{2.12} describes 
a dot in the absence of the spin-orbit interaction, which would destroy 
this symmetry. 

Electron transport through the dot occurs via two dot-lead junctions.
In a typical geometry, the confining potential forming a lateral quantum 
dot varies smoothly on the scale of the Fermi wavelength. 
Hence, the point contacts connecting the quantum dot to the leads act 
essentially as electronic waveguides. Potentials on the gates control 
the waveguide width, and, therefore, the number of electronic modes 
the waveguide support: by making the waveguide narrower one pinches 
the propagating modes off one-by-one. Each such mode contributes 
$2e^2/h$ to the conductance of a contact. The Coulomb blockade 
develops when the conductances of the contacts are small compared 
to $2e^2/h$, i.e. when the very last propagating mode approaches 
its pinch-off~\cite{KM,KF}.  Accordingly, in the Coulomb blockade 
regime each dot-lead junction in a lateral quantum dot system supports 
only a single electronic mode~\cite{real}.

As discussed below, for $E_C\gg\delta E$ the characteristic energy 
scale relevant to the Kondo effect, the Kondo temperature $T_K$, is 
small compared to the mean level spacing: $T_K\ll\delta E$. This 
separation of the energy scales allows us to simplify the problem 
even further by assuming that the conductances of the dot-lead 
junctions are small. This assumption will not affect the properties 
of the system in the Kondo regime. At the same time, it justifies 
the use of the tunneling Hamiltonian for description of the coupling 
between the dot and the leads. The microscopic Hamiltonian of the 
system can then be written as a sum of three distinct terms,
\beq
H = H_{\rm leads} + H_{\rm dot}+ H_{\rm tunneling},
\label{2.14}
\eeq
which describe free electrons in the leads, isolated quantum dot, 
and tunneling between the dot and the leads, respectively. The 
second term in~\Eq{2.14}, the Hamiltonian of the dot 
$H_{\rm dot}$, is given by \Eq{2.12}. We treat the leads as 
reservoirs of free electrons with continuous spectra $\xi_{k}$, 
characterized by constant density of states $\nu$, same for both 
leads. Moreover, since the typical energies $\omega\lesssim E_C$ 
of electrons participating in transport through a quantum dot in 
the Coulomb blockade regime are small compared to the Fermi 
energy of the electron gas in the leads, the spectra $\xi_{k}$ can 
be linearized near the Fermi level, $\xi_k = v_F k$; here $k$ is 
measured from $k_F$. With only one electronic mode per junction 
taken into account, the first and the third terms in \Eq{2.14} have the form
\begin{eqnarray}
&& H_{\rm leads} =  \sum_{\alpha ks}\xi^\pdag_k 
c^\dagger_{\alpha ks} c^\pdag_{\alpha ks} ,
\quad
\xi_k = - \xi_{-k},
\label{2.15} \\
&& H_{\rm tunneling} 
= \sum_{\alpha k n s}t^\pdag_{\alpha n^\pprime}\!
c^\dagger_{\alpha k s} d^\pdag_{ns} + {\rm H.c.}
\label{2.16}
\end{eqnarray}
Here $t_{\alpha n}$ are tunneling matrix elements (tunneling 
amplitudes) ``connecting'' the state $n$ in the dot with the state 
$k$ in the lead $\alpha$ ($\alpha =R,L$ for the right/left lead).

Tunneling leads to a broadening of discrete levels in the dot. 
The width $\Gamma_{\alpha n}$ that level $n$ acquires due to 
escape of an electron to lead $\alpha$ is given by
\beq
\Gamma_{\alpha n} = \pi\nu\left|t^2_{\alpha n}\right|
\label{2.17}
\eeq 
Randomness of single-particle states in the dot translates into 
the randomness of the tunneling amplitudes. Indeed, the amplitudes 
depend on the values of the electron wave functions at the points 
$\bm r_\alpha$ of the contacts, $t_{\alpha n}\propto \phi_n(\bm r_\alpha)$. 
For $k_F \left|\bm r_L - \bm r_R\right|\sim k_F L\gg 1$ the tunneling 
amplitudes [and, therefore, the widths \eqref{2.17}] in the left and right 
junctions are statistically independent of each other. Moreover, the 
amplitudes to different energy levels are also uncorrelated, see \Eq{2.6a}:
\beq
\overline{\,t^\ast_{\alpha n^\pprime}\! t^\past_{\alpha' n '}} 
= \frac{\Gamma_\alpha}{\pi\nu} \,
\delta^\past_{\alpha\alpha'} \delta^\past_{nn'},
\quad
\overline{\,t^\past_{\alpha n^\pprime}\! t^\past_{\alpha' n '}}
= \frac{\Gamma_\alpha}{\pi\nu} \,
\delta^\past_{\beta,1}\delta^\past_{\alpha\alpha'} \delta^\past_{nn'},
\label{2.18}
\eeq
The average value 
$\Gamma_{\alpha} = \overline{\,\Gamma_{\alpha n}}$ 
of the width is related to the conductance of the corresponding 
dot-lead junction
\beq
G_{\alpha} = \frac{4e^2}{\hbar} \frac{\Gamma_\alpha}{\delta E}\,. 
\label{2.19}
\eeq
In the regime of strong Coulomb blockade $(G_\alpha\ll e^2/h)$,
the widths are small compared to the level spacing, 
$\Gamma_\alpha\ll \delta E$, so that discrete levels in the dot are 
well defined. Note that statistical fluctuations of the widths 
$\Gamma_{\alpha n}$ are large, and the corresponding distribution 
function is not Gaussian. Indeed, using Eqs.~\eqref{2.17} and 
\eqref{2.18} it is straightforward~\cite{Haake,Efetov} to show that
\beq
P(\gamma) = \overline{\,\delta\left(\gamma 
- {\Gamma_{\alpha n}}/{\Gamma_\alpha}\right)\,}
= \left\{
\begin{array}{cr}
\dfrac{e^{-\gamma/2}}{\sqrt{2\pi\gamma}}\,, & \beta = 1
\\ \\
e^{-\gamma}, &\beta = 2
\end{array}
\right.
\label{2.20}
\eeq
This expression is known as Porter-Thomas distribution~\cite{PT}.

\section{Thermally-activated conduction}
\label{activation}

At high temperatures, $T\gg E_C $, charging energy is negligible 
compared to the thermal energy of electrons. Therefore the 
conductance of the device in this regime $G_\infty$ is not 
affected by charging and, independently of the gate voltage, 
is given by
\beq
\frac{1}{\,G_\infty}=\frac{1}{G_L}+\frac{1}{G_R}.
\label{3.1}
\eeq
Dependence on $N_0$ develops at lower temperatures, 
$T\lesssim E_C$. It turns out that the conductance is suppressed 
for all gate voltages except narrow regions (\textit{Coulomb blockade 
peaks}) around half-integer values of $N_0$. We will demonstrate 
this now using the method of rate equations~\cite{rate,rate_discrete}. 

\subsection{Onset of the Coulomb blockade oscillations}
\label{peak_high_T}

We start with the regime of relatively high temperatures,
\beq
\delta E\ll T\ll E_C, 
\label{3.2}
\eeq
and assume that the gate voltage is tuned sufficiently close 
to one of the points of charge degeneracy,
\beq
|N_0^\past - N_0^*|\lesssim T/E_C 
\label{3.3}
\eeq
(here $N_0^*$ is a half-integer number). 

Condition~\eqref{3.2} enables us to treat the discrete 
single-particle levels within the dot as a continuum with the density 
of states $1/\delta E$.
Condition~\Eq{3.3}, on the other hand, allows us to take into 
account only two charge states of the dot which are almost 
degenerate in the vicinity of the Coulomb blockade peak. For 
$N_0^\past$ close to $N_0^*$ these are the state $\ket{0}$ 
with $N = N_0^*- 1/2$ electrons on the dot, and the state 
$\ket{1}$ with $N = N_0^*+ 1/2$ electrons. According to 
Eqs.~\eqref{2.12} and \eqref{3.3}, the difference of electrostatic 
energies of these states (the energy cost to add an electron to the 
dot) is
\beq
E_+(N_0) = E_{\ket{1}}-E_{\ket{0}} 
= 2E_C (N_0^*-N_0) \lesssim T.
\label{3.4a}
\eeq

In addition to the constraints~\eqref{3.2} and \eqref{3.3}, we 
assume here that the inelastic electron relaxation rate within the 
dot $1/\tau_\epsilon$ is large compared to the escape rates 
$\Gamma_\alpha/\hbar$. In other words, transitions between 
discrete levels in the dot occur before the electron escapes to 
the leads\footnote{Note that a finite inelastic relaxation rate 
requires inclusion of mechanisms beyond the model~\Eq{2.12}, 
e.g., electron-phonon collisions.}. 
Under this assumption the tunnelings across the two junctions 
can be treated independently of each other (this is known as 
\textit{sequential tunneling} approximation). 

With the help of the Fermi golden rule the current $I_\alpha$ from 
the lead $\alpha$ into the dot can be written as
\beqa
&&I_{\alpha} = e \,\frac{2\pi}{\hbar}
\sum_{kns} \left|t_{\alpha n}^2\right| 
\delta(\xi_k + eV_\alpha - \epsilon_n -E_{+}) 
\label{3.4b}\\ 
&&\qquad\qquad\qquad
\times\Bigl\{{\cal P}_0 f(\xi_k)[1-f(\epsilon_n)] 
- {\cal P}_1 f(\epsilon_n)[1-f(\xi_k)]\Bigr\}\,.
\nonumber
\eeqa
Here ${\cal P}_i$ is the probability to find the dot in the charge 
states $\ket{i}$ $(i=0,1)$, $f(\omega) = [\exp(\omega/T)+1]^{-1}$ 
is the Fermi function, and $V_\alpha$ is the electric potential on 
the lead $\alpha$, see Fig.~\ref{circuit}. 
In writing \Eq{3.4b} we assumed that the distribution functions 
$f(\xi_k)$ and $f(\epsilon_n)$ are not perturbed. This is well 
justified provided that the relaxation rate $1/\tau_\epsilon$ 
exceeds the rate $\sim G_\infty|V_L-V_R|/e$ at which electrons 
pass through the dot. Replacing the summations over $n$ and $k$
in \Eq{3.4b} by integrations over the corresponding continua, and 
making use of Eqs.~\eqref{2.17} and \eqref{2.19}, we find 
\beq
I_\alpha = \frac{G_\alpha}{e}\bigl[{\cal P}_0 F(E_{+} - eV_\alpha)
- {\cal P}_1 F(eV_\alpha -E_{+})\bigr],
\quad
F(\omega)=\frac{\omega}{e^{\omega/T}-1}\,.
\label{3.4}
\eeq
In the steady state, the currents across the two junctions satisfy
\beq
I = I_L = -I_R .
\label{3.5}
\eeq
Equations~\eqref{3.4} and~\eqref{3.5}, supplemented by the obvious
normalization condition ${\cal P}_0 + {\cal P}_1 =1$, 
allow one to find the probabilities ${\cal P}_{i}$ and the current across
the dot $I$ in response to the applied bias $V = V_L - V_R$. 
This yields for the linear conductance across the dot~\cite{rate}
\beq
G =\lim_{V\to 0}dI/dV = G_\infty
\frac{E_C (N_0^\past - N_0^*)/T}{\sinh [2E_C (N_0^\past - N_0^*)/T]} \,.
\label{3.6}
\eeq

Here $N_0^\past = N_0^*$ corresponds to the Coulomb blockade 
peak. At each peak, the conductance equals half of its high-temperature 
value $G_\infty$, see \Eq{3.1}. On the contrary, in the 
\textit{Coulomb blockade valleys} ($N_0^\past \neq N_0^*$), the 
conductance falls off exponentially with the decrease of temperature, 
and all the valleys behave exactly the same way.  Note that the 
sequential tunneling approximation disregards any interference 
phenomena for electrons passing the dot. Accordingly, the result 
\Eq{3.6} is insensitive to a weak magnetic field.

\subsection{Coulomb blockade peaks at low temperature}
\label{peak_low_T}

At temperatures below the single-particle level spacing in the dot
$\delta E$, the activation energy for electron transport equals the
difference between the ground state energies of the
Hamiltonian~\eqref{2.12} corresponding to two subsequent (integer)
eigenvalues of $N$. Obviously, this difference includes, in addition
to the electrostatic contribution $E_+(N_0)$, see \Eq{3.4a}, also a
finite (and random) level spacing.  As a result, the distance in $N_0$
between adjacent Coulomb blockade peaks is no longer $1$, but contains
a small fluctuating contribution of the order of $\delta E/E_C$.
Mesoscopic fluctuations of spacings between the peaks are still
subject of a significant disagreement between theory and experiments.
We will not consider these fluctuations here (see~\cite{Alhassid_RMP}
for a recent review), and discuss only the heights of the peaks.

We concentrate on the temperature interval
\beq
\Gamma_\alpha\ll T\ll\delta E,
\label{3.7}
\eeq 
which extends to lower temperatures the regime considered in the
previous section, see \Eq{3.2}, and on the gate voltages tuned to 
the vicinity of the Coulomb blockade peak, see~\Eq{3.3}. Just as 
above, the latter condition allows us to neglect all charge states 
except the two with the lowest energy, $\ket{0}$ and $\ket{1}$. 
Due to the second inequality in \Eq{3.7}, the thermal broadening 
of single-particle energy levels in the dot can be neglected, and the 
states $\ket{0}$ and $\ket{1}$ coincide with the ground states of 
the Hamiltonian~\eqref{2.12} with, respectively, $N = N_0^*-1/2$ 
and $N = N_0^*+1/2$ electrons in the dot. 
To be definite, consider the case when
\beq
N_0^* = N + 1/2
\label{3.8a}
\eeq
with $N$ being an even integer; for simplicity, we also neglect the
exchange term in \Eq{2.12}. Then $\ket{0}$ (with even number 
of electrons $N$) is the state in which all single particle levels 
below the Fermi level $(n<0)$ are doubly occupied. This state is, 
obviously, non-degenerate. The state $\ket{1}$ differs from $\ket{0}$ 
by an addition of a single electron on the Fermi level $n=0$. The extra 
electron may be in two possible spin states, hence $\ket{1}$ is 
doubly degenerate; we denote the two components of $\ket{1}$ 
by $\ket{s}$ with $s=\uparrow,\downarrow$.  As discussed below, 
the degeneracy eventually gives rise to the Kondo effect. However, 
at $T\gg \Gamma_\alpha$ [see \Eq{3.7}] the quantum coherence 
associated with the onset of the Kondo effect is not important, and 
the rate equations approach can still be used to study the transport 
across the dot~\cite{rate_discrete}. 

Applying the Fermi golden rule, we write the contribution of electrons 
with spin $s$ to the electric current $I_{\alpha s}$ from lead $\alpha$ 
to the dot as
\[
I_{\alpha s} = e \,\frac{2\pi}{\hbar}\left|t_{\alpha 0}^2\right| 
\sum_{k} \delta\left(\xi_k + eV_\alpha - \epsilon_0 - E_{+}\right) 
\Bigl\{{\cal P}_0 f(\xi_k) - {\cal P}_s \left[1-f(\xi_k)\right]\Bigr\}
\]
We now neglect $\epsilon_0$ as it is small compared to $E_+$ 
(thereby neglecting the mesoscopic fluctuations of the position of 
the Coulomb blockade peak) and replace the summation over $k$ 
by an integration. This yields
\beq
I_{\alpha s} = \frac{2e}{\hbar}\,\Gamma_{\alpha 0} 
\Bigl\{{\cal P}_0 f\left(E_{+}-eV_\alpha\right) 
- {\cal P}_s f\left(eV_\alpha-E_{+}\right) \Bigr\}.
\label{3.8}
\eeq
In the steady state the currents $I_{\alpha s}$ satisfy
\beq
I_{Ls} = -I_{Rs} = I/2
\label{3.9}
\eeq
(here we took into account that both projections of spin 
contribute equally to the total electric current across the dot $I$).
Solution of Eqs.~\eqref{3.8} and \eqref{3.9} subject to the 
normalization condition 
${\cal P}_0 + {\cal P}_\uparrow + {\cal P}_\downarrow = 1$
results in~\cite{ABG}
\beq
G = \frac{4e^2}{\hbar}
\frac{\Gamma_{L0}\Gamma_{R0}}{\Gamma_{L0}+\Gamma_{R0}}\,
\left[\negthinspace \frac{-df/d\omega}{~~~1+f(\omega)}\right]_{\omega= E_+(N_0)}.
\label{3.10}
\eeq
The case of odd $N$ in \Eq{3.8a} is also described by \Eq{3.10} 
after replacement $E_+(N_0)\to - E_+(N_0)$. 

There are several differences between \Eq{3.10} and the corresponding 
expression \Eq{3.6} valid in the temperature range~\eqref{3.2}. First 
of all, the maximum of the conductance \Eq{3.10} occurs at the gate 
voltage slightly (by an amount of the order of $T/E_C$) off the 
degeneracy point $N_0=N_0^*$, and, more importantly, the shape 
of the peak is not symmetric about the maximum. This asymmetry is 
due to correlations in transport of electrons with opposite spins through 
a single discrete level in the dot. In the maximum, the function~\eqref{3.10}
takes value
\beq
G_{\rm peak} \sim 
\frac{e^2}{h}
\frac{\Gamma_{L0}\Gamma_{R0}}{\Gamma_{L0}+\Gamma_{R0}}\,
\frac{1}{T}\,.
\label{3.11}
\eeq

Note that Eqs.~\eqref{3.10} and \eqref{3.11} depend on the widths 
$\Gamma_{\alpha 0}$ of the energy level $n=0$ rather then on the 
averages $\Gamma_\alpha$ over many levels in the dot, as in \Eq{3.6}. 
As already discussed in Sec.~\ref{model}, the widths $\Gamma_{\alpha 0}$ 
are related to the values of the electron wave functions at the position 
of the dot-lead contacts, and, therefore, are random. Accordingly, the 
heights $G_{\rm peak}$ of the Coulomb blockade peaks exhibit 
strong mesoscopic fluctuations. In view of \Eq{2.20}, the distribution 
function of $G_{\rm peak}$, see \Eq{3.11}, is expected to be broad 
and strongly non-Gaussian, as well as very sensitive to the magnetic 
flux threading the dot. This is indeed confirmed by 
calculations~\cite{JSA,PEI} and agrees with experimental 
data~\cite{Folk96,Chang96}. The expression for the distribution 
function is rather cumbersome and we will not reproduce it here, 
referring the reader to the original papers~\cite{JSA,PEI} and  
reviews~\cite{Alhassid_RMP,Efetov,ABG}) instead. 

An order-of-magnitude estimate of the average height of 
the peak can be obtained by replacing $\Gamma_{\alpha 0}$ in 
\Eq{3.11} by $\Gamma_\alpha$, see \Eq{2.19}, which yields 
\beq
\overline{\,G_{\rm peak}}
\sim G_\infty \frac{\delta E}{T}\,.
\label{3.12}
\eeq
This is by a factor $\delta E/T$ larger than the corresponding 
figure $G_{\rm peak} = G_\infty/2$ for the temperature range~\eqref{3.2}, 
and may even approach the unitary limit $(\sim e^2/h)$ at the lower 
end of the temperature interval~\eqref{3.7}. Interestingly, breaking 
of time-reversal symmetry results in an increase of the average 
conductance~\cite{ABG}. This increase is analogous to negative 
magnetoresistance due to weak localization in bulk 
systems~\cite{magnetoresistance}, with the same physics involved.

\section{Activationless transport through a blockaded quantum dot}

According to the rate equations theory~\cite{rate}, at low 
temperatures, $T\ll E_C $, conduction through the dot is 
exponentially suppressed in the Coulomb blockade valleys. 
This suppression occurs because the process of electron 
transport through the dot involves a \textit{real transition} 
to the state in which the charge of the dot differs by $e$ from 
the thermodynamically most probable value. The probability 
of such fluctuation is proportional to 
$\exp\left(-E_C |N_0^\past - N_0^*|/T\right)$, 
which explains the conductance suppression, see \Eq{3.5}. 
Going beyond the lowest-order perturbation theory in 
conductances of the dot-leads junctions $G_\alpha$ allows 
one to consider processes in which states of the dot with a 
``wrong'' charge participate in the tunneling process as 
\textit{virtual states}. The existence of these higher-order 
contributions to the tunneling conductance was envisioned 
already in 1968 by Giaever and Zeller~\cite{Giaever}. The 
first quantitative theory of this effect, however, was developed 
much later~\cite{AN}. 

The leading contributions to the activationless transport, 
according to~\cite{AN}, are provided by the processes 
of \textit{inelastic and elastic co-tunneling}. Unlike the sequential 
tunneling, in the co-tunneling mechanism, the events of electron 
tunneling from one of the leads into the dot, and tunneling from 
the dot to the other lead occur as a single quantum process. 

\subsection{Inelastic co-tunneling}

\begin{figure}[h]
\centerline{\includegraphics[width=0.95\textwidth]{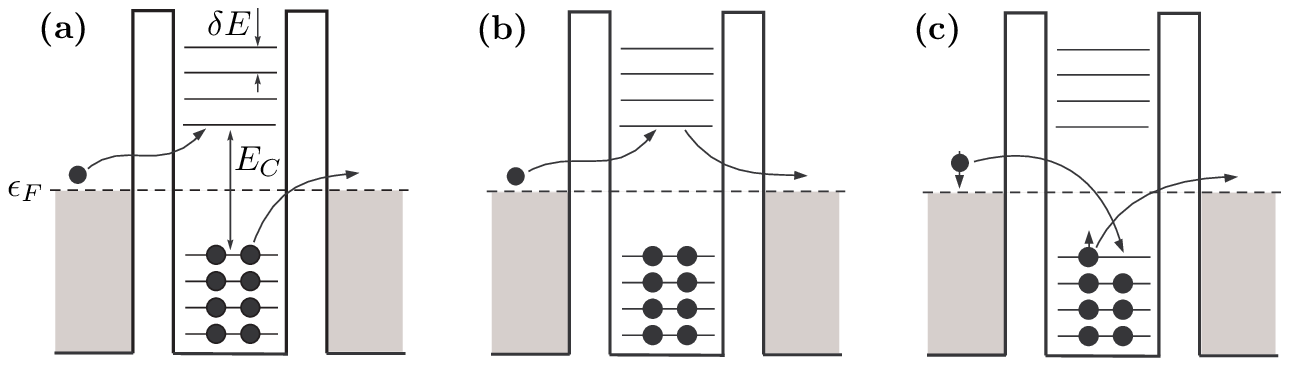}}
\vspace{2mm}
\caption{Examples of the co-tunneling processes. 
\newline (a) inelastic co-tunneling: transferring of an electron 
between the leads leaves behind an electron-hole pair in the dot;
(b) elastic co-tunneling;
(c) elastic co-tunneling with a flip of spin.
}
\label{cotunneling}
\end{figure}

In the inelastic co-tunneling mechanism, an electron tunnels 
from a lead into one of the vacant single-particle levels in 
the dot, while it is an electron occupying some other level 
that tunnels out of the dot, see Fig.~\ref{cotunneling}(a). As 
a result, transfer of charge $e$ between the leads is accompanied 
by a simultaneous creation of an electron-hole pair in the dot.

Here we will estimate the contribution of the inelastic co-tunneling 
to the conductance deep in the Coulomb blockade valley, i.e. 
at almost integer $N_0$. Consider an electron that tunnels into 
the dot from the lead $L$. If energy $\omega$ of the electron 
relative to the Fermi level is small compared to the charging energy, 
$\omega\ll E_C $, then the energy of the virtual state involved in the
co-tunneling process is close to $E_C $. The amplitude of the 
inelastic transition via this virtual state to the lead $R$ is then 
given by
\beq
A_{n,m} 
= \frac{t^\ast_{L n^\pprime} {\negthinspace} t^\past_{R m}}{E_C }.
\label{4.1}
\eeq
The initial state of this transition has an extra electron in the
single-particle state $k$ in the lead $L$, while the final state has 
an extra electron in the state $k'$ in the lead $R$ and an 
electron-hole pair in the dot (state $n$ is occupied, state 
$m$ is empty). The conductance is proportional to the sum 
of probabilities of all such processes, 
\beq
G_{in} \sim \frac{e^2}{h} \sum_{n,m} \nu^2 \left|A_{n,m}^2\right|
\sim \frac{e^2}{h} \frac{1}{E_C^2} 
\sum_{n,m} \Gamma_{Ln}\Gamma_{Rm},
\label{4.1a}
\eeq
where we made use of \Eq{2.17}. Now we estimate how many terms 
contribute to the sum in \eqref{4.1a}.
Given the energy of the initial state $\omega$, the number of available 
final states can be found from the phase space argument, familiar 
from the calculation of the quasiparticle lifetime in the Fermi liquid 
theory~\cite{Abrikosov}. For $\omega\gg\delta E$ the energies of the 
states $n$ and $m$ lie within a strip of the width $\sim \omega/\delta E$ 
about the Fermi level. The total number of the final states contributing to 
the sum in \Eq{4.1a} is then of the order of $(\omega/\delta E)^2$. 
Since the typical value of $\omega$ is $T$, the average value of the 
inelastic co-tunneling contribution to the conductance can be estimated as 
\[
\overline{\,G_{in}} 
\sim \left(\frac{T}{\delta E}\right)^2 \frac{e^2}{h} 
\frac{\Gamma_L\Gamma_R}{E_C^2} .
\]
Using now \Eq{2.19}, we find~\cite{AN}
\beq
\overline{\,G_{in}} 
\sim \frac{G_L G_R}{e^2/h} \left(\frac{T}{E_C }\right)^2.
\label{4.2}
\eeq

The terms entering the sum in \Eq{4.1a} are random and uncorrelated. 
As it follows from \Eq{2.20}, fluctuation of each term in the sum is of the 
order of its average value. This leads to the estimate for the fluctuation 
$\delta G_{in} = G_{in}-\overline{G_{in}}$ of the inelastic co-tunneling 
contribution to the conductance
\beq
\overline{\,\delta G_{in}^2} 
\sim \left(\frac{T}{\delta E}\right)^2 
\left(\frac{e^2}{h} \frac{\Gamma_L\Gamma_R}{E_C^2}\right)^2
\sim \left(\frac{\delta E}{T}\right)^2\bigl(\overline{G_{in}}\,\bigr)^2 \,.
\label{4.2a}
\eeq
Accordingly, fluctuation of $G_{in}$ is small compared to its average.

A comparison of \Eq{4.2} with the result of the rate equations
theory \eqref{3.6} shows that the inelastic co-tunneling takes over
the thermally-activated hopping at moderately low temperatures
\beq
T\lesssim T_{\rm in} =
E_C \left[\ln\left(\frac{e^2/h}{G_L+G_R}\right)\right]^{-1}.
\label{4.3}
\eeq
The smallest energy of an electron-hole pair is of the order of 
$\delta E$. At temperatures below this threshold the inelastic 
co-tunneling contribution is exponentially suppressed. It turns 
out, however, that even at much higher temperatures this 
mechanism becomes less effective than the elastic co-tunneling.

\subsection{Elastic co-tunneling}

In the process of elastic co-tunneling, transfer of charge between 
the leads is not accompanied by the creation of an electron-hole pair 
in the dot. In other words, occupation numbers of single-particle 
energy levels in the dot in the initial and final states of the co-tunneling 
process are exactly the same, see Fig.~\ref{cotunneling}(b). Close to 
the middle of the Coulomb blockade valley (at almost integer $N_0$)
the average number of electrons on the dot, $N\approx N_0$, is also 
an integer. Both an addition and a removal of a single electron cost 
$E_C$ in electrostatic energy, see \Eq{2.12}. The amplitude of the 
elastic co-tunneling process in which an electron is transfered from 
lead $L$ to lead $R$ can then be written as
\beq
A_{el} = \sum_n 
t^\ast_{Ln} t^\past_{Rn}
\frac{\,\sign(\epsilon_n)}{E_C + |\epsilon_n|}
\label{4.4}
\eeq
The two types of contributions to the amplitude $A_{el}$ are
associated with virtual creation of either an electron if the level $n$ 
is empty ($\epsilon_n > 0$), or of a hole if the level is occupied 
($\epsilon_n < 0$); the relative sign difference between the two 
types of contributions originates in the fermionic commutation 
relations.

As discussed in Sec.~\ref{model}, the tunneling amplitudes 
$t_{\alpha n}$ entering \Eq{4.4} are Gaussian random variables 
with zero mean and variances given by \Eq{2.18}. It is then easy 
to see that the second moment of the amplitude~\Eq{4.4} is given by
\[
\overline{\,\left|A^2_{el}\right|}
= \frac{\Gamma_L\Gamma_R}{\,(\pi\nu)^2}
\sum_n \bigl(E_C + |\epsilon_n|\bigr)^{-2}.
\]
Since for $E_C\gg \delta E$ the number of terms making significant 
contribution to the sum over $n$ here is large, and since the sum is 
converging, one can replace the summation by an integral which 
yields
\beq
\overline{\,\left|A^2_{el}\right|}
\approx \frac{\Gamma_L\Gamma_R}{\,(\pi\nu)^2}
\frac{1}{E_C\delta E}\,.
\label{4.5}
\eeq
Substitution of this expression into
\beq
G_{el} = \frac{4\pi e^2 \nu^2}{\hbar} \left|A^2_{el}\right|
\label{4.6}
\eeq
and making use of \Eq{2.19} gives~\cite{AN} 
\beq
\overline{\,G_{el}} 
\sim \frac{G_L G_R}{e^2/h}  \frac{\delta E}{E_C }\,.
\label{4.7}
\eeq
for the average value of the elastic co-tunneling contribution to 
the conductance. 

This result is easily generalized to gate voltages tuned away 
from the middle of the Coulomb blockade valley. The 
corresponding expression reads 
\beq
\overline{\,G_{el}} 
\sim \frac{G_L G_R}{e^2/h}  \frac{\delta E}{E_C }
\left(\frac{1}{N_0-N_0^*}+\frac{1}{N_0^*-N_0+1}\right)\,.
\label{4.8}
\eeq
and is valid when $N_0$ is not too close to the degeneracy points
$N_0 = N_0^*$ and $N_0=N_0^*+1$ ($N_0^*$ is a half-integer number): 
\[
\min\bigl\{
\left|N_0-N_0^*\right|,\left|N_0-N_0^*-1\right|
\bigr\} \gg \delta E/E_C
\]

Comparison of \Eq{4.7} with \Eq{4.2} shows that the elastic 
co-tunneling mechanism dominates the electron transport 
already at temperatures
\beq
T\lesssim T_{el} = \sqrt{E_C \delta E} \,,
\label{4.9}
\eeq
which may exceed significantly the level spacing. However, as we 
will see shortly below, mesoscopic fluctuations of $G_{el}$ are 
strong~\cite{AG}, of the order of its average value. Thus, although 
$\overline{G_{el}}$ is always positive, see \Eq{4.8}, the 
sample-specific value of $G_{el}$ for a given gate voltage may 
vanish.

The key to understanding the statistical properties of the elastic 
co-tunneling contribution to the conductance is provided by the
observation that there are many $(\sim E_C/\delta E\gg 1)$
terms making significant contribution to the amplitude \Eq{4.4}.
All these terms are random and statistically independent of each other.
The central limit theorem then suggests that the distribution of $A_{el}$
is Gaussian~\cite{ABG}, and, therefore, is completely characterised 
by the first two statistical moments,
\beq
\overline{\,A^\past_{el}} = \overline{\,A^\ast_{el}},
\quad
\overline{\,A^\past_{el}A^\past_{el}} 
= \overline{\,A^\ast_{el}A^\ast_{el}} 
= \delta^\past_{\beta,1}\overline{\,A^\past_{el}A^\ast_{el}}
\label{4.10}
\eeq
with $\overline{A^\ast_{el}A^\past_{el}}$ given by \Eq{4.5}. 
This can be proven by explicit consideration of higher moments. 
For example, 
\beq
\overline{\,\left|A^4_{el}\right|}
= 2\Bigl(\overline{\,A^\past_{el}A^\ast_{el}}\Bigr)^2 
+ \Bigl|\overline{\,A^\past_{el}A^\past_{el}}\,\Bigr|^2 +
\delta \overline{\,\left|A^4_{el}\right|}.
\label{4.11}
\eeq
The non-Gaussian correction here, 
$
\delta \overline{\,\left|A^4_{el}\right|} 
\sim \overline{\,\left|A^2_{el}\right|}\,(\delta E/E_C)
$, 
is by a factor of $\delta E/E_C\ll 1$ smaller than the main 
(Gaussian) contribution.

It follows from Eqs.~\eqref{4.6}, \eqref{4.10}, and \eqref{4.11} 
that the fluctuation of the conductance 
$\delta G_{el} = G_{el}-\overline{G_{el}}$ satisfies
\beq
\overline{\,\delta G_{el}^2} = \frac{2}{\beta} 
\bigl(\overline{G_{el}}\,\bigr)^2 \,.
\label{4.12}
\eeq
Note that breaking of time reversal symmetry reduces the fluctuations 
by a factor of 2, similar to conductance fluctuations in bulk systems, 
whereas the average conductance \eqref{4.8} is not affected by the 
magnetic field. 

It is clear from \Eq{4.12} that the fluctuations of the conductance 
are of the order of the conductance itself, despite naive expectations 
that the large number of the contributing states results in self-averaging. 
The reason is that one has to add amplitudes, rather than probabilities, 
in order to compute the conductance. Because the fluctuations of the 
conductance are large, its distribution function is not Gaussian. Given 
the statistics \eqref{4.10} of the amplitudes, it is not quite surprising 
that the distribution of $G_{el}$ normalized to its average coincides 
with Porter-Thomas distribution \eqref{2.20}. This result was obtained 
first in~\cite{AG} by a different (and more general) method.

Finally, it is interesting to compare the elastic co-tunneling contribution 
to the conductance fluctuations \Eq{4.12} with that of inelastic 
co-tunneling, see \Eq{4.2a}. Even though the inelastic co-tunneling 
is the main conduction mechanism at $T\gtrsim T_{el}$, see \Eq{4.9},
elastic co-tunneling dominates the fluctuations of the conductance 
throughout the Coulomb blockade regime $T\lesssim E_C$.
 
\section{Kondo regime in transport through a quantum dot}
\label{Kondo}

In the above discussion of the elastic co-tunneling we made a tacit 
assumption that all single-particle levels in the dot are either empty 
or doubly occupied. This, however, is not the case when the dot 
has a non-zero spin in the ground state. A dot with odd number 
of electrons, for example, would necessarily have a half-integer 
spin $S$. In the most important case of $S=1/2$ the top-most 
occupied single-particle level is filled by a single electron and is 
spin-degenerate. This opens a possibility of a co-tunneling process 
in which a transfer of an electron between the leads is accompanied 
by a flip of electron's spin with simultaneous flip of the spin on the 
dot, see Fig.~\ref{cotunneling}(c). 

The amplitude of such a process, calculated in the fourth order 
in tunneling matrix elements, diverges logarithmically when the 
energy $\omega$ of an incoming electron approaches $0$. Since 
$\omega\sim T$, the logarithmic singularity in the transmission 
amplitude translates into a dramatic enhancement of the 
conductance $G$ across the dot at low temperatures: $G$ may 
reach values as high as the quantum limit $2e^2/h$~\cite{AM, unitary}. 
This conductance enhancement is not really a surprise. Indeed, in 
the spin-flip co-tunneling process a quantum dot with odd $N$ 
behaves as $S=1/2$ magnetic impurity embedded into a tunneling 
barrier separating two massive conductors~\cite{Erice}. It is 
known~\cite{old_reviews} since mid-60's that the presence 
of such impurities leads to zero-bias anomalies in tunneling 
conductance~\cite{classics}, which are adequately 
explained~\cite{Appelbaum,Anderson} in the context of the 
Kondo effect~\cite{Kondo}. 

\subsection{Effective low-energy Hamiltonian}
\label{H_K}

At energies well below the threshold $\Delta\sim\delta E$ for 
intra-dot excitations the transitions within the $(2S+1)$-fold 
degenerate ground state manifold of a dot can be conveniently 
described by a spin operator ${\bf S}$. The form of the 
\textit{effective Hamiltonian} describing the interaction of the 
dot with conduction electrons in the leads is then dictated by 
SU(2) symmetry\footnote{In writing \Eq{5.1} we omitted the 
potential scattering terms associated with the usual elastic 
co-tunneling. This approximation is well justified when the 
conductances of the dot-lead junctions are small, 
$G_\alpha \ll e^2/h$, in which case $G_{el}$ is also very small, 
see \Eq{4.5}},
\beq
H_{\rm ef\mbox{}f} = \sum_{\alpha ks}\xi^\pdag_{k} 
c^\dagger_{\alpha ks} c^\pdag_{\alpha ks}
+ \sum_{\alpha \alpha'} J_{\alpha\alpha'}
({\bf s}_{\alpha'\alpha} \cdot {\bf S}) 
\label{5.1} 
\eeq
with ${\bf s}_{\alpha \alpha'} = \sum_{kk'ss'}
c_{\alpha ks}^{\dagger} (\bm\sigma_{ss'}/2) \,c_{\alpha' k's'}^\pdag$. 
The sum over $k$ in \Eq{5.1} is restricted to $|\xi_k|<\Delta$. 
The exchange amplitudes $J^\past_{\alpha\alpha'}$ form 
$2\times 2$ Hermitian matrix $\hat J$. The matrix has two 
real eigenvalues, the exchange constants $J_1$ and $J_2$ 
(hereafter we assume that $J_1\geq J_2$). By an appropriate 
rotation in the $R-L$ space the Hamiltonian~\eqref{5.2} can 
then be brought into the form
\beq
H_{\rm ef\mbox{}f} = \sum_{\gamma ks} \xi^\pdag_{k} 
\psi^\dagger_{\gamma ks} \psi^\pdag_{\gamma ks}
+ \sum_{\gamma} J_\gamma
({\bf s}_\gamma\cdot{\bf S}).
\label{5.2}
\eeq
Here the operators $\psi_\gamma$ are certain linear combinations 
of the original operators $c_{R,L}$ describing electrons in the 
leads, and 
\[
{\bf s}_\gamma 
= \sum_{kk'ss'} \psi_{\gamma ks}^{\dagger} \frac{\bm\sigma_{ss'}}{2} 
\,\psi_{\gamma k's'}^\pdag
\] 
is local spin density of itinerant electrons in the ``channel" 
$\gamma =1,2$. 

The symmetry alone is not sufficient to determine the exchange 
constants $J_\gamma$;  their evaluation must rely upon a microscopic 
model. Here we briefly outline the derivation~\cite{real,Fiete,JPCM} 
of \Eq{5.1} for a generic model of a quantum dot system discussed 
in Section~\ref{model} above. For simplicity, we will assume that 
the gate voltage $N_0$ is tuned to the middle of the Coulomb 
blockade valley.  The tunneling~\eqref{2.16} mixes the state with 
$N = N_0$ electrons on the dot with states having $N\pm 1$ 
electrons. The electrostatic energies of these states are high 
$(\sim E_C )$, hence the transitions $N\to N\pm 1$ are virtual, 
and can be taken into account perturbatively in the second order 
in tunneling amplitudes~\cite{SW}. 

For the Hamiltonian~\eqref{2.12} the occupations of single-particle 
energy levels are good quantum numbers. Therefore, the amplitude 
$J_{\alpha\alpha'}$ can be written as a sum of partial amplitudes,
\beq
J_{\alpha\alpha'} = \sum_n J_{\alpha\alpha'}^n .
\label{5.3}
\eeq
Each term in the sum here corresponds to a process during which 
an electron or a hole is created virtually on the level $n$ in the dot, 
cf. \Eq{4.4}. For $G_\alpha \ll e^2/h$ and $E_S \ll \delta E$ the 
main contribution to the sum in~\Eq{5.3} comes from 
singly-occupied energy levels in the dot. A dot with spin $S$ has 
$2S$ such levels near the Fermi level (hereafter we assign indexes 
$n= -S,\ldots,n=S$ to these levels), each carrying a spin 
${\bf S}/2S$, and contributing 
\beq
J_{\alpha\alpha'}^n = \frac{\lambda_n}{E_C}\,
t^\ast_{\alpha n}  t^\past_{\alpha' n},
\quad 
\lambda_n = 2/S,
\quad
|n|\leq S
\label{5.4}
\eeq
to the exchange amplitude in~\Eq{5.1}. This yields
\beq
J_{\alpha\alpha'}  
\approx \sum_{|n|\leq S} J_{\alpha\alpha'}^n .
\label{5.5}
\eeq
It follows from Equations~\eqref{5.3} and~\eqref{5.4} that 
\beq
{\rm tr} \hat J = \frac{1}{E_C} \sum_n \lambda_n 
\left(|t^2_{Ln}| + |t^2_{Rn}|\right) .
\label{5.6}
\eeq
By restricting the sum over $n$ here to $|n|\leq S$, as in~\Eq{5.5}, 
and taking into account that all $\lambda_n$ in~\Eq{5.4} are positive, 
we find $J_1 + J_2 > 0$. Similarly, from
\beq
\det\hat J = \frac{1}{2E^2_C} \sum_{m,n} \lambda_m \lambda_n 
|\mathcal D_{mn}^2|,
\quad
\mathcal D_{mn} = \det 
\left(
\begin{array}{lr}
t_{Lm}& t_{Rm} 
\\
t_{Ln}& t_{Rn} 
\end{array}
\right)
\label{5.7}
\eeq
and Equations~\eqref{5.4} and~\eqref{5.5} follows that $J_1 J_2 > 0$ 
for $S>1/2$. Indeed, in this case the sum in~\Eq{5.7} contains at least 
one contribution with $m\neq n$; all such contributions are positive.
Thus, both exchange constants $J_{1,2} > 0$ if the dot's spin 
$S$ exceeds $1/2$~\cite{real}. The peculiarities of the Kondo effect 
in quantum dots with large spin are discussed in Section~\ref{large_S} 
below. 

We now turn to the most common case of $S=1/2$ on the 
dot~\cite{kondo_exp}. The ground state of such dot has only 
one singly-occupied energy level $(n=0)$, so that $\det\hat J \approx 0$, 
see~Eqs.~\eqref{5.5} and \eqref{5.7}.  Accordingly, one of the 
exchange constants vanishes,
\beq
J_2 \approx 0,
\label{5.8}
\eeq
while the remaining one, $J_1 = {\rm tr}\hat J$, is positive. 
Equation~\eqref{5.8} resulted, of course, from the approximation 
made in~\Eq{5.5}. For the model~\eqref{2.12} the leading correction 
to~\Eq{5.5} originates in the co-tunneling processes with an intermediate 
state containing an extra electron (or an extra hole) on one of the empty 
(doubly-occupied) levels. Such contribution arises because the spin 
on the level $n$ is not conserved by the Hamiltonian~\eqref{2.12}, 
unlike the corresponding occupation number. Straightforward 
calculation~\cite{Fiete} yields the partial amplitude in the form 
of \Eq{5.4}, but with
\[
\lambda_n = - \frac{2E_C E_S}{(E_C + |\epsilon_n|)^2},
\quad
\quad n\neq 0 .
\]

Unless the tunneling amplitudes $t_{\alpha 0}$ to the only 
singly-occupied level in the dot are anomalously small, the 
corresponding correction 
\beq
\delta J_{\alpha\alpha'} = \sum_{n\neq 0} J_{\alpha\alpha'}^n 
\label{5.9}
\eeq
to the exchange amplitude~\eqref{5.5} is small,
\[
\left|\frac{\delta J_{\alpha\alpha'}}
{J_{\alpha\alpha'}}\right| \sim \frac{E_S}{\delta E} \ll 1,
\]
see \Eq{2.13}. To obtain this estimate, we assumed that all 
tunneling amplitudes $t_{\alpha n}$ are of the same order of 
magnitude, and replaced the sum over $n$ in~\Eq{5.9} by 
an integral. A similar estimate yields the leading contribution 
to $\det\hat J$,
\[
\det \hat J \approx \frac{1}{E^2_C} \sum_n \lambda_0 \lambda_n 
|\mathcal D_{0n}^2|
\sim - \frac{E_S}{\delta E} \bigl({\rm tr}\hat J\bigr)^2,
\]
or
\beq
J_2/J_1 \sim - E_S/\delta E .
\label{5.10}
\eeq

According to~\Eq{5.10}, the exchange constant $J_2$ is
negative~\cite{SI}, and its absolute value is small compared to $J_1$.
Hence, \Eq{5.8} is indeed an excellent approximation for large
chaotic dots with spin $S=1/2$ as long as the intra-dot exchange
interaction remains weak,\footnote{Equation~\eqref{5.8} 
holds identically for the Anderson impurity model~\cite{Anderson}
frequently employed to study transport through quantum 
dots. In that model a quantum dot is described by a single energy 
level, which formally corresponds to the infinite level spacing limit 
$\delta E\to\infty$ of the Hamiltonian~\eqref{2.12}.} 
i.e. for $E_S \ll \delta E$.  
Note that corrections to the universal Hamiltonian~\eqref{2.12} also 
result in finite values of both exchange constants, $|J_2|\sim J_1 N^{-1/2}$, 
and become important for small dots with $N\lesssim 10$~\cite{unitary}. 
Although this may significantly affect the conductance across the system 
in the weak coupling regime $T\gtrsim T_K$, it does not lead to qualitative 
changes in the results for $S = 1/2$ on the dot, as the channel with 
smaller exchange constant decouples at low energies~\cite{NB}, see 
also Section~\ref{large_S} below. With this caveat, we adopt the 
approximation~\eqref{5.8} in our description of the Kondo effect in 
quantum dots with spin $S=1/2$. Accordingly, the effective Hamiltonian 
of the system~\eqref{5.2} assumes the``block-diagonal'' form
\beqa
&&H_{\rm ef\mbox{}f} = H_1 + H_2 
\label{5.11} \\
&&H_1 = \sum_{ks}\xi^\pdag_{k} \psi^\dagger_{1 ks} \psi^\pdag_{1 ks}
+ J ({\bf s}_1 \cdot {\bf S}) 
\label{5.12} \\
&&H_2 = \sum_{ks}\xi^\pdag_{k} \psi^\dagger_{2ks} \psi^\pdag_{2ks}
\label{5.13} 
\eeqa
with $J = {\rm tr} \hat J > 0$. 


To get an idea about the physics of the Kondo model 
(see~\cite{Kondo_reviews} for recent reviews), let us first 
replace the fermion field operator ${\bf s}_1$ in \Eq{5.12} 
by a single-particle spin-1/2 operator ${\bf S}_1$. The ground 
state of the resulting Hamiltonian of two spins 
\[
\widetilde{H} = J({\bf S}_1\cdot{\bf S})
\] 
is obviously a singlet. The excited state (a triplet) is separated 
from the ground state by the energy gap $J_1$. This separation 
can be interpreted as the binding energy of the singlet. Unlike 
${\bf S}_1$ in this simple example, the operator ${\bf s}_1$ 
in~\Eq{5.12} is merely a spin density of the conduction electrons 
at the site of the ``magnetic impurity". Because conduction electrons 
are freely moving in space, it is hard for the impurity to ``capture" 
an electron and form a singlet. Yet, even a weak local exchange 
interaction suffices to form a singlet ground 
state~\cite{PWA_book,Wilson}. 
However, the characteristic energy (an analogue of the binding 
energy) for this singlet is given not by the exchange constant $J$, 
but by the so-called Kondo temperature
\beq
T_K \sim \Delta \exp(-1/\nu J).
\label{6.1}
\eeq
Using $\Delta\sim\delta E$ and Equations~\eqref{5.6} and~\eqref{2.19}, 
one obtains from~\Eq{6.1} the estimate
\beq
\ln\left(\frac{\delta E}{T_K}\right)
\sim \frac{1}{\nu J}
\sim \frac{e^2/h}{G_L + G_R}\frac{E_C }{\delta E}.
\label{6.2}
\eeq
Since $G_\alpha\ll e^2/h$ and $E_C \gg \delta E$, the r.h.s. 
of~\Eq{6.2} is a product of two large parameters. Therefore, 
the Kondo temperature $T_K $ is small compared to the mean 
level spacing,
\beq
T_K \ll \delta E.
\label{6.3}
\eeq
It is this separation of the energy scales that justifies the use of the 
effective low-energy Hamiltonian~\eqref{5.1},~\eqref{5.2} for the 
description of the Kondo effect in a quantum dot system. 
The inequality~\eqref{6.3} remains valid even if the conductances 
of the dot-leads junctions $G_\alpha$ are of the order of $2e^2/h$. 
However, in this case the estimate~\eqref{6.2} is no longer
applicable~\cite{GHL}. 

In our model, see Equations~\eqref{5.11}-\eqref{5.13}, one of 
the channels $(\psi_2)$ of conduction electrons completely 
decouples from the dot, while the $\psi_1$-particles are described 
by the standard single-channel antiferromagnetic Kondo 
model~\cite{Kondo,Kondo_reviews}. Therefore, the 
thermodynamic properties of a quantum dot in the Kondo regime 
are identical to those of the conventional Kondo problem for a 
single magnetic impurity in a bulk metal; thermodynamics of the 
latter model is fully studied~\cite{bethe}.  However, all the 
experiments addressing the Kondo effect in quantum dots test 
their transport properties rather then thermodynamics. The electron 
current operator is not diagonal in the $(\psi_1,\psi_2)$ representation, 
and the contributions of these two sub-systems to the conductance 
are not additive. Below we relate the linear conductance and, in some 
special case, the non-linear differential conductance as well, to the 
t-matrix of the conventional Kondo problem.

\subsection{Linear response}

The linear conductance can be calculated from the Kubo formula
\beq
G = \lim_{\omega\to 0} \frac{1}{\hbar\,\omega}
\int_0^\infty d t \,e^{i\omega t} 
\Bigl\langle \bigl[\hat I(t), \hat I(0) \bigr]\Bigr\rangle ,
\label{6.4} 
\eeq
where the current operator $\hat I$ is given by
\beq
\hat I = \frac{d}{dt} \frac{e}{2} 
\bigl(\hat N_{R} - \hat N_{L}\bigr), 
\quad
\hat N_\alpha 
= \sum_{ks}^\pdag c^\dagger_{\alpha ks}c^\pdag_{\alpha ks}
\label{6.5} 
\eeq
Here $\hat N_\alpha$ is the operator of the total number of electrons 
in the lead $\alpha$. Evaluation of the linear conductance proceeds 
similarly to the calculation of the impurity contribution to the resistivity 
of dilute magnetic alloys (see, e.g.,~\cite{AL}). In order to take the 
full advantage of the decomposition~\eqref{5.11}-\eqref{5.13}, we rewrite 
$\hat I$ in terms of the operators $\psi_{1,2}$. These operators are 
related to the original operators $c_{R,L}$ representing the electrons 
in the right and left leads via
\beq
\left(\begin{array}{c}
\psi_{1ks} \\ 
\psi_{2ks}
\end{array}\right)
= \left(\begin{array}{cc}
\cos\theta_0 & \sin\theta_0
\\ 
-\sin\theta_0 & \cos\theta_0 
\end{array}\right)
\left(\begin{array}{c}
c_{Rks} \\ 
c_{Lks}
\end{array}\right) .
\label{6.6} 
\eeq
The rotation matrix here is the same one that diagonalizes matrix
$\hat J$ of the exchange amplitudes in~\Eq{5.1}; the rotation angle
$\theta_0$ satisfies the equation $\tan \theta_0 = |t_{L0}/t_{R0}|$.
With the help of \Eq{6.6} we obtain
\beq
\hat N_R - \hat N_L = \cos(2\theta_0) \bigl(\hat N_1 - \hat N_2\bigr) 
- \sin(2\theta_0) \sum_{ks}
\bigl(
\psi_{1ks}^\dagger \psi_{2ks}^\pdag + 
{\rm H.c.}
\bigr) 
\label{6.7} 
\eeq
The current operator $\hat I$ entering the Kubo formula~\eqref{6.4} 
is to be calculated with the equilibrium Hamiltonian~\eqref{5.11}-\eqref{5.13}. 
Since both $\hat N_1$ and $\hat N_2$ commute with $H_{\rm ef\mbox{}f}$, 
the first term in~\Eq{6.7} makes no contribution to $\hat I$. When 
the second term in~\Eq{6.7} is substituted into~\Eq{6.5} and then 
into the Kubo formula~\eqref{6.4}, the result, after integration by parts, 
can be expressed via 2-particle correlation functions such as 
$\bigl\langle \psi^\dagger_1(t)\psi_2^\pdag(t) 
\psi_2^\dagger(0)\psi_1^\pdag(0)\bigr\rangle$ 
(see Appendix B of~\cite{PG} for further details about this calculation). 
Due to the block-diagonal structure of $H_{\rm ef\mbox{}f}$, 
see~\Eq{5.11}, these correlation functions factorize into products 
of the single-particle correlation functions describing the (free) 
$\psi_2$-particles and the (interacting) $\psi_1$-particles. The result 
of the evaluation of the Kubo formula can then be written as 
\beq
G = G_0 \int d\omega 
\left(-\frac{df}{d\omega}\right) 
\frac{1}{2} \sum_s \bigl[- \pi\nu \im T_s (\omega)\bigr]. 
\label{6.8}
\eeq
Here 
\beq
G_0 = \frac{2e^2}{h}  \sin^2(2\theta_0) 
= \frac{2e^2}{h} 
\frac{4|t^2_{L0} t^2_{R0}|}{\bigl(|t_{L0}^2|+|t_{R0}^2|\bigr)^2} \,,
\label{6.9}
\eeq
\noindent 
$f(\omega)$ is the Fermi function, and $T_s(\omega)$ is the 
t-matrix for the Kondo model~\eqref{5.12}. The t-matrix is 
related to the exact retarded Green function of the 
$\psi_1$-particles in the conventional way,
\[
G_{ks,k'\negthinspace s}(\omega) 
= G^0_k(\omega) + G^0_k (\omega)T_s (\omega) G^0_{k'}(\omega), 
\quad
G^0_k = (\omega - \xi_k +i0)^{-1} .
\]
Here $G_{ks,k'\negthinspace s}(\omega)$ is the Fourier transform of 
$G_{ks,k'\negthinspace s}(t) = -i\theta(t)
\bigl\langle\bigl\{
\psi_{1ks}^\pdag(t),\psi_{1k'\negthinspace s}^\dagger
\bigr\}\bigr\rangle$, 
where $\langle\ldots\rangle$ stands for the thermodynamic averaging 
with the Hamiltonian~\eqref{5.12}. In writing \Eq{6.8} we took into 
account the conservation of the total spin (which implies that 
$G_{ks,k'\negthinspace s'}=\delta_{ss'}G_{ks,k'\negthinspace s}$, 
and that the interaction in~\Eq{5.12} is local (which in turn means 
that the t-matrix is independent of $k$ and $k'$). 

\subsection{Weak coupling regime: $T_K\ll T\ll \delta E$}
\label{weak_coupling}

When the exchange term 
in the Hamiltonian~\eqref{5.12} is treated perturbatively, the main 
contribution to the t-matrix comes from the transitions of the type~\cite{AAA}
\beq
\ket{ks,\sigma}\to\ket{k'\negthinspace s'\negthinspace ,\sigma'}.
\label{1}
\eeq
Here the state $\ket{ks,\sigma}$ has an extra electron 
with spin $s$ in the orbital state $k$ whereas the dot is in the spin state 
$\sigma$. By SU(2) symmetry, the amplitude of the transition~\eqref{1}
satisfies
\beq
A_{\ket{k'\negthinspace s'\negthinspace ,\sigma'}\leftarrow \ket{ks,\sigma}} 
= A(\omega) \frac{1}{4} 
\left(\bf\sigma_{s'\negthinspace s}\cdot\bf\sigma_{\sigma'\negthinspace \sigma}\right)
\label{2}
\eeq
Note that the amplitude is independent of $k,k'$ because the interaction 
is local. However, it may depend on $\omega$ due to retardation effects. 

The transition~\eqref{1} is \textit{elastic} in the sense that the
number of quasiparticles in the final state of the transition is the
same as that in the initial state (in other words, the transition~\eqref{1} 
is not accompanied by the production of electron-hole pairs). Therefore, 
the imaginary part of the t-matrix can be calculated with the help of the 
optical theorem~\cite{Newton}, which yields
\beq
-\pi\nu \im T_s 
= \frac{1}{2}\sum_\sigma \sum_{s'\negthinspace\sigma'}
\left| \pi\nu \,
A_{\ket{k'\negthinspace s'\negthinspace ,\sigma'}
\leftarrow \ket{ks,\sigma}}^2
\right|\,.
\label{3}
\eeq
The factor $1/2$ here accounts for the probability to have spin 
$\sigma$ on the dot in the initial state of the transition. Substitution 
of the tunneling amplitude in the form~\eqref{2} into \Eq{3}, and 
summation over the spin indexes with the help of the identity~\eqref{0} 
result in
\beq
-\pi\nu \im T_s = \frac{3\pi^2}{16\,} \,\nu^2\! \left|A^2(\omega)\right| .
\label{4}
\eeq
In the leading (first) order in $J$ one readily obtains $A^{(1)} = J$, 
independently of $\omega$. However, as discovered by Kondo~\cite{Kondo},
the second-order contribution $A^{(2)}$ not only depends on $\omega$,
but is logarithmically divergent as $\omega\to 0$:
\[
A^{(2)}(\omega) = \nu J^2 \ln\left|\Delta/\omega\right|. 
\]
Here $\Delta$ is the high-energy cutoff in the Hamiltonian~\eqref{5.12}.
It turns out~\cite{AAA} that similar logarithmically divergent contributions
appear in all orders of perturbation theory, 
\[
\nu A^{(n)} (\omega) = (\nu J)^n \bigl[\ln|\Delta/\omega|\bigr]^{n-1},
\]
resulting in a geometric series
\[
\nu A(\omega) = \sum_{n=1}^{\infty} \nu A^{(n)} 
= \nu J \sum_{n=0}^{\infty}\bigl[\nu J \ln|\Delta/\omega|\bigr]^n
=  \frac{\nu J}{1 - \nu J\ln|\Delta/\omega|} \,.
\]
With the help of~\Eq{6.1} this can be written as
\beq
\nu A(\omega) = \frac{1}{\ln|\omega/T_K|}\,.
\label{5}
\eeq
Substitution of~\Eq{5} into \Eq{4} and then into \Eq{6.8}, 
and evaluation of the integral over $\omega$ with logarithmic 
accuracy yield for the conductance across the dot
\beq
G = G_0 \frac{3\pi^2/16}{\ln^2(T/T_K)}\,, 
\quad
T\gg T_K .
\label{6.12}
\eeq
Equation \eqref{6.12} is the leading term of the asymptotic expansion in 
powers of $1/\ln(T/T_K)$, and represents the conductance 
in the {\it leading logarithmic approximation}. 

Equation \Eq{6.12} resulted from summing up the most-diverging 
contributions in all orders of perturbation theory. It is instructive 
to re-derive it now in the framework of
\textit{renormalization group}~\cite{PWA}. The idea of this 
approach rests on the observation that the electronic states that
give a significant contribution to observable quantities, such as 
conductance, are states within an interval of energies of the 
width $\omega\sim T$ about the Fermi level, see \Eq{6.8}. 
At temperatures of the order of $T_K$, when the Kondo effect 
becomes important, this interval is narrow compared to the width
of the band $D=\Delta$ in which the Hamiltonian~\eqref{5.12} is 
defined.

Consider a narrow strip of energies of the width $\delta D\ll D$ near 
the edge of the band. Any transition~\eqref{1} between a state near the 
Fermi level and one of the states in the strip is associated with high 
($\sim D$) energy deficit, and, therefore, can only occur virtually.
 Obviously, virtual transitions via each of the states in the strip result 
in the second-order correction $\sim J^2/D$ to the amplitude $A(\omega)$ 
of the transition between states in the vicinity of the Fermi level. Since 
the strip contains $\nu \delta D$ electronic states, the total correction 
is~\cite{PWA} 
\[
\delta A \sim \nu J^2\delta D/D .
\] 
This correction can be accounted for by modifying the exchange
constant in the effective Hamiltonian $\widetilde{H}_{\rm ef\mbox{}f}$
which is defined for states within a narrower energy band of the width
$D-\delta D$~\cite{PWA},
\begin{eqnarray}
&& \widetilde{H}_{\rm ef\mbox{}f}
= 
\sum_{ks} \xi^\pdag_k 
{\psi}^\dagger_{ks}{\psi}^\pdag_{ks}
+ J_{D - \delta D}
({\bf s}_\psi \cdot{\bf S}),
\quad
|\xi_k|<D-\delta D,
\label{6} \\
&& J_{D - \delta D} = J_D + \nu J_D^2 \frac{\delta D}{D}.
\label{7}
\end{eqnarray}
Here $J_D$ is the exchange constant in the original Hamiltonian. 
Note that the $\widetilde{H}_{\rm ef\mbox{}f}$ has the same 
form as \Eq{5.12}. This is not merely a conjecture, but can be 
shown rigorously~\cite{Wilson,AYH}. 

The reduction of the bandwidth can be considered to be a result 
of a unitary transformation that decouples the states near the band 
edges from the rest of the band~\cite{flow}. In principle, any such 
transformation should also affect the operators that describe the 
observable quantities. Fortunately, this is not the case for the problem 
at hand. Indeed, the angle $\theta_0$ in \Eq{6.6} is not modified by 
the transformation. Therefore, the current operator and the expression 
for the conductance~\eqref{6.8} retain their form.

Successive reductions of the high-energy cutoff $D$ by small steps 
$\delta D$ can be viewed as a continuous process during which the 
initial Hamiltonian~\eqref{5.12} with $D=\Delta$ is transformed to an 
effective Hamiltonian of the same form that acts within the band of the 
reduced width $D\ll \Delta $. It follows from~\Eq{7} that the dependence 
of the effective exchange constant on $D$ is described by the differential 
equation~\cite{PWA,AYH}
\beq
\frac{dJ_D}{d\zeta} =\nu J_D^2,
\quad
\zeta = \ln\left({\Delta}/{D}\right).
\label{8}
\eeq
With the help of \Eq{6.1}, the solution of the RG equation~\eqref{8} 
subject to the initial condition $J_\Delta = J$ can be cast into the form
\beq
\nu J_D = \frac{1}{\ln(D/T_K)}\,.
\label{8a}
\eeq
The renormalization described by \Eq{8} can be continued until 
the bandwidth $D$ becomes of the order of the typical energy 
$|\omega|\sim T$ for real transitions. After this limit has been reached, 
the transition amplitude $A(\omega)$ is calculated in lowest (first) 
order of perturbation theory in the effective exchange constant (higher 
order contributions are negligible for $D\sim \omega$ ),
\[
\nu A(\omega) = \nu J_{D\sim|\omega|} = \frac{1}{\ln|\omega/T_K|}
\]
Using now Eqs.~\eqref{4} and~\eqref{6.8}, we recover \Eq{6.12}.

\subsection{Strong coupling regime: $T\ll T_K$}
\label{strong_coupling}

As temperature approaches $T_K$, the leading logarithmic 
approximation result \Eq{6.12} diverges. This divergence signals 
the failure of the approximation. Indeed, we are considering a model 
with single-mode junctions between the dot and the leads. The 
maximal possible conductance in this case is $2e^2/h$. To obtain 
a more precise bound, we discuss in this section the conductance 
in the strong coupling regime $T\ll T_K$.

We start with the zero-temperature limit $T=0$. As discussed above, 
the ground state of the Kondo model~\eqref{5.12} is a singlet~\cite{PWA_book}, 
and, accordingly, is not degenerate. Therefore, the t-matrix of the conduction 
electrons interacting with the localized spin is completely characterized 
by the scattering phase shifts $\delta_s$ for electrons with spin $s$ 
at the Fermi level. The t-matrix is then given by the standard 
scattering theory expression~\cite{Newton}
\beq
-\pi\nu \,T_s(0) = \frac{1}{2i}\left(\mathbb{S}_s -1\right) ,
\quad \mathbb{S}_s = e^{2i\delta_s},
\label{6.13}
\eeq
where $\mathbb{S}_s$ is the scattering matrix for electrons with 
spin $s$, which for a single channel case reduces to its eigenvalue. 
Substitution of~\Eq{6.13} into \Eq{6.8} yields
\beq
G(0) = G_0\frac{1}{2} \sum_s \sin^2\delta_s
\label{6.14}
\eeq
for the conductance, see \Eq{6.8}. The phase shifts
in Eqs.~\eqref{6.13},~\eqref{6.14} are obviously defined only 
${\rm mod}\,\pi$ (that is, $\delta_s$ and $\delta_s +\pi$ are 
equivalent). This ambiguity can be removed if we set to zero the 
values of the phase shifts at $J = 0$ in \Eq{5.12}.

In order to find the two phase shifts $\delta_s$, we need two 
independent relations. The first one follows from the invariance 
of the Kondo Hamiltonian~\eqref{5.12} under the particle-hole 
transformation $\psi^\pdag_{ks} \to s\psi^\dagger_{-k,-s}$ (here 
$s=\pm 1$ for spin-up/down electrons). The particle-hole symmetry 
implies the relation for the t-matrix
\beq
T_s(\omega) = - T^*_{-s}(-\omega),
\label{6.15}
\eeq
valid at all $\omega$ and $T$. In view of 
\Eq{6.13}, it translates into the relation for the phase shifts 
at the Fermi level ($\omega = 0$)~\cite{N},
\beq
\delta_\uparrow + \delta_\downarrow = 0.
\label{6.16}
\eeq

The second relation follows from the requirement that the 
ground state of the Hamiltonian~\eqref{5.12} is a singlet~\cite{N}. 
In the absence of exchange ($J=0$) and at $T=0$, an 
infinitesimally weak (with Zeeman energy $B\to + 0$) magnetic 
field 
\beq
H_B = BS^z
\label{6.17}
\eeq 
would polarize the dot's spin. Since free electron gas has zero 
spin in the ground state, the total spin in a very large but finite 
region of space $\cal V$ surrounding the dot coincides with the 
spin of the dot, $\langle S^z\rangle_{J=0} = -1/2$. If the exchange 
with electron gas is now turned on, $J > 0$, a very weak field will 
not prevent the formation of a singlet ground state. In this state, 
the total spin within $\cal V$ is zero. Such change of the spin is 
possible if the numbers of spin-up and spin-down electrons within 
$\cal V$ have changed to compensate for the dot's spin: 
$\delta N_\uparrow -\delta N_\downarrow = 1$.
By the Friedel sum rule, $\delta N_s$ are related to the scattering phase 
shifts at the Fermi level, $\delta N_s = \delta_s/\pi $, which gives
\beq
\delta_\uparrow - \delta_\downarrow = \pi.
\label{6.18}
\eeq
Combining Eqs.~\eqref{6.16} and~\eqref{6.18}, we find 
\beq
\delta_s =s \frac{\pi}{2}\,.
\label{6.18a}
\eeq
Equation \eqref{6.14} then yields for zero-temperature and zero-field
conductance across the dot~\cite{AM}
\beq
G(0) = G_0.
\label{6.19}
\eeq
Thus, the grow of the conductance with lowering the temperature 
is limited only by the value of $G_0$. This value, see \Eq{6.9}, 
depends only on the ratio of the tunneling amplitudes $|t_{L0}/t_{R0}|$;  
if $|t_{L0}|=|t_{R0}|$, then the conductance at $T=0$ will reach 
the maximal value $G=2e^2/h$ allowed by quantum mechanics~\cite{AM}.

As explained above, screening of the dot's spin by itinerant electrons 
amounts to $\pm\pi/2$ phase shifts for electrons at the Fermi level 
$(\omega=0)$. 
However, the phase shifts at a finite $\omega$ deviate from the 
unitary limit value. Such deviation is to be expected, as the Kondo 
``resonance'' has a finite width $\sim T_K$. There is also an inelastic
component of scattering appearing at finite $\omega$. 
Virtual transitions to excited states, corresponding to a broken up 
Kondo singlet, induce a local repulsive interaction between itinerant 
electrons~\cite{N}, which is the cause of inelastic scattering.
This interaction enters the fixed-point Hamiltonian, applicable 
when the relevant energies are small compared to the Kondo 
temperature $T_K$. The fixed point Hamiltonian takes a relatively 
simple form~\cite{N} when written in the basis of electronic 
states that incorporate an extra $\pm\pi/2$ phase shift 
[see \Eq{6.18a}] compared to the original basis of \Eq{5.12}.
In the new basis,   
\beq
H_{\text{fixed point}} 
= \sum_{ks}\xi_k \varphi^\dagger_{ks} \varphi^\pdag_{ks}
-\sum_{kk'\!s} 
\frac{\xi_k+\xi_{k'}\!}{2\pi\nu T_K}\,
\varphi^\dagger_{ks} \varphi^\pdag_{k'\!s}
+ \,\frac{1}{\pi\nu^2 T_K}\, 
\rho_{\uparrow}\rho_{\downarrow}.
\label{6.100}
\eeq
Here 
$\rho_s = \sum_{kk'}\!\!:\!\!\varphi^\dagger_{ks} \varphi^\pdag_{k'\!s}\!\!:$ 
(the colons denote the normal ordering). 
The form of the last two terms in the r.h.s. of \Eq{6.100} is dictated 
by the particle-hole symmetry~\cite{N}. The ratio of their coefficients 
is fixed by the physical requirement that the Kondo singularity is 
tied to the Fermi level~\cite{N} (i.e. that the phase shifts depend 
only on the distance $\omega$ to the Fermi level), while the overall 
coefficient $1/T_K$ can be viewed as a precise definition of the 
Kondo temperature\footnote{With this convention the RG 
expression~\Eq{6.1} is regarded as an estimate of $T_K$ 
with logarithmic accuracy.}. Note that by construction, the t-matrix 
for $\varphi-$particles $\widetilde {\,T}$ vanishes for $\omega =T= 0$. 

The second term in the r.h.s. of \Eq{6.100} describes a 
purely elastic scattering, and yields a small $\omega-$dependent 
phase shift~\cite{N}
\beq
\widetilde\delta(\omega) = \frac{\omega}{T_K}\,.
\label{6.20a}
\eeq
The last term in \Eq{6.100} gives rise to an inelastic contribution to 
$\widetilde {\,T}$. Evaluation of this contribution in the second 
order of perturbation theory (see~\cite{AL,N} for details) results in
\beq
-\pi\nu \widetilde {\,T}_{in}(\omega) 
= i\,\frac{\omega^2 +\pi^2T^2}{2\,T_K^2}\,.
\label{6.20b}
\eeq

In order to obtain the t-matrix in the original basis $T_s(\omega)$, 
we note that the elastic scattering phase shifts here are obtained 
by simply adding $\pm \pi/2$ to $\widetilde\delta$, i.e.
\beq
\delta_s(\omega) =  s\frac{\pi}{2} + \widetilde\delta(\omega).
\label{6.20c}
\eeq
The relation between $T_s(\omega)$ and $\widetilde {\,T}(\omega)$,
accounting for small inelastic term $\widetilde {\,T}_{in}$, reads 
\beq
-\pi\nu T_s(\omega) 
= \frac{1}{2i}\left[e^{2i\delta_s(\omega)} -1\right] + e^{2i\delta_s(\omega)}
\left[-\pi\nu \widetilde {\,T}_{in}(\omega)\right]. 
\label{6.20d}
\eeq
Note that the inelastic contribution enters this expression with an extra 
factor of $\exp(2i\delta_s)$. The appearance of this factor is a direct 
consequence of the unitarity of the scattering matrix in the presence 
of both elastic and inelastic scattering channels, see~\cite{N}.
Taking into account Eqs.~\eqref{6.20a}, \eqref{6.20b}, and 
\eqref{6.20c}, we find
\beq
-\pi\nu \im T_s (\omega) 
= 1-\frac{3\omega^2 +\pi^2T^2}{2\,T_K^2}\,.
\label{6.20}
\eeq
Substitution of this expression into \Eq{6.8} then yields
\beq
G = G_0\left[1 - \left({\pi T}/{T_K}\right)^2\right],
\quad
T\ll T_K\,.
\label{6.21}
\eeq
Accordingly, corrections to the conductance are quadratic in 
temperature -- a typical Fermi liquid result~\cite{N}. The 
weak-coupling ($T\gg T_K$) and the strong-coupling ($T\ll T_K$)
asymptotes of the conductance have a strikingly different structure.
Nevertheless, since the Kondo effect is a crossover phenomenon 
rather than a phase transition~\cite{Kondo_reviews,PWA_book,Wilson,bethe}, 
the dependence $G(T)$ is a smooth and featureless~\cite{Costi} 
function throughout the crossover region $T\sim T_K$.

\begin{figure}[h]
\centerline{\includegraphics[width=0.7\textwidth]{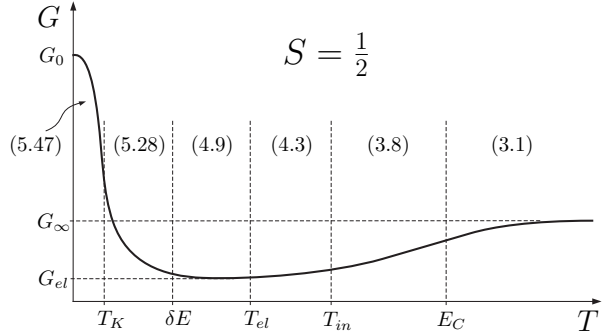}}
\vspace{2mm}
\caption{Sketch of the temperature dependence of the conductance 
in the middle of the Coulomb blockade valley with $S = 1/2$ on the 
dot. The numbers in brackets refer to the corresponding equations 
in the text.
}
\label{overall}
\end{figure}

Finally, note that both Eqs.~\eqref{6.12} and \eqref{6.21} have
been obtained here for the particle-hole symmetric model~\eqref{5.12}. 
This approximation is equivalent to neglecting the elastic co-tunneling 
contribution to the conductance $G_{el}$. The 
asymptotes~\eqref{6.12},~\eqref{6.21} remain valid~\cite{real} as long 
as $G_{el}/ G_0\ll 1$. The overall temperature dependence of the 
linear conductance in the middle of the Coulomb blockade valley 
is sketched in Fig.~\ref{overall}.

The fixed point Hamiltonian~\eqref{6.100} also allows one to 
calculate the corrections to zero-temperature conductance due to 
a finite magnetic field, which enters \Eq{6.100} via a term
$H_B = \sum_{ks}(s/2)B\,\varphi^\dagger_{ks} \varphi^\pdag_{ks}$.
The field polarizes the electron gas, and the operators $\rho_s$ acquire 
non-zero ground state expectation value 
$\langle\rho_s\rangle = -\nu B s/2$. In order to calculate the phase shifts 
at the Fermi energy we replace $\rho_\uparrow \rho_\downarrow$ 
in the third term in the r.h.s. \eqref{6.100} by 
$\sum_s \langle\rho_{-s}\rangle \rho_s$, and $\xi_k, \xi_{k'}$ in the 
second term by their values at the Fermi level $\xi_k = - Bs/2$. 
As a result, \Eq{6.100} simplifies to
\[
H = \sum_{ks}(\xi_k+\frac{s}{2}B) \varphi^\dagger_{ks} \varphi^\pdag_{ks}
+\sum_{kk'\!s} 
\frac{Bs}{\pi\nu T_K}\,\rho_s,
\]
from which one can read off the scattering phase shifts for $\varphi-$particles,
\[
\widetilde\delta_s = - sB/T_K.
\]
The phase shifts in the original basis are given by~\cite{N} 
\[
\delta_s = s\frac{\pi}{2} + \widetilde\delta_s 
= s\left(\frac{\pi}{2}-\frac{B}{T_K}\right),
\]
cf. \Eq{6.20c}. Equation \eqref{6.14} then gives
\beq
G = G_0\left[1 - \left(B/T_K\right)^2\right],
\quad
B\ll T_K\,.
\label{6.21b}
\eeq

Note that the effect of a finite magnetic field on zero-temperature 
conductance is very similar to that a finite temperature has on the 
conductance at zero field, see \Eq{6.21}. The similarity is not limited 
to the strong coupling regime. Indeed, the counterpart of \Eq{6.12}
reads
\beq
G = G_0 \frac{\pi^2/16}{\ln^2(B/T_K)}\,, 
\quad
B\gg T_K .
\label{6.21c}
\eeq

\subsection{Beyond linear response}
\label{NONEQ}

In order to study transport through a quantum dot away from 
equilibrium we add to the effective Hamiltonian~\eqref{5.11}-\eqref{5.13} 
a term 
\beq
H_V = \frac{eV}{2} \left(\hat N_L - \hat N_R\right)
\label{6.22}
\eeq
describing a finite voltage bias $V$ applied between the left and right 
electrodes. Here we will evaluate the current across the dot at arbitrary 
$V$ but under the simplifying assumption that the dot-lead junctions are 
strongly asymmetric: 
\[
G_L\ll G_R.
\] 
Under this condition the angle $\theta_0$ in~\Eq{6.6} is small, 
$\theta_0\approx |t_{L0}/t_{R0}|\ll 1$. Expanding \Eq{6.7} 
to linear order in $\theta_0$ we obtain
\beq
H_V(\theta_0) = \frac{eV}{2} \bigl(\hat N_2- \hat N_1\bigr) 
+ eV \theta_0 \sum_{ks}
\bigl( \psi_{1ks}^\dagger \psi_{2ks}^\pdag + {\rm H.c.} \bigr)
\label{6.23} 
\eeq
The first term in the r.h.s. here can be interpreted as the voltage 
bias between the reservoirs of $1$- and $2$-particles, cf.~\Eq{6.22}, 
while the second term has an appearance of $k$-conserving tunneling 
with very small (proportional to $\theta_0\ll 1$) tunneling amplitude. 

Similar to \Eq{6.23}, the current operator $\hat I$, see~\Eq{6.5}, 
splits naturally into two parts,
\begin{eqnarray*}
&&\hat I = \hat I_0 + \delta \hat I, 
\\
&& \hat I_0 = \frac{d}{dt} \frac{e}{2} 
\bigl(\hat N_1- \hat N_2\bigr) 
= - i e^2 V \theta_0 \sum_{ks} 
\psi_{1ks}^\dagger \psi_{2ks}^\pdag + {\rm H.c.} ,
\\
&&\delta\hat I = - e\,\theta_0 \frac{d}{dt} \sum_{ks} 
\psi_{1ks}^\dagger \psi_{2ks}^\pdag + {\rm H.c.}
\end{eqnarray*}
It turns out that $\delta\hat I$ does not contribute to the average 
current in the leading (second) order in $\theta_0$~\cite{Erice}. 
Indeed, in this order
\[
\langle\delta \hat I\rangle=i\theta_0^2e^2V\frac{d}{dt}\int_{-\infty}^t dt'
\left\langle[\hat{\cal O}(t),\hat{\cal O}(t')]\right\rangle_0,
\quad 
\hat{\cal O}=\sum_{ks}\psi^\dagger_{1ks}\psi_{2ks},
\]
where $\langle\dots\rangle_0$ denotes thermodynamic averaging with
the Hamiltonian $H_{\rm eff}+H_V(0)$. The thermodynamic (equilibrium)
averaging is well defined here, because for $\theta_0 = 0$ the Hamiltonian 
$H_{\rm eff}+H_V$ conserves separately the numbers of $1$- and
$2$-particles. Taking into account that
$\langle[\hat{\cal O}(t),\hat{\cal O}(t')]\rangle_0$ depends only on the
difference $\tau=t-t'$, we find
\[
\langle\delta \hat I\rangle=i\theta_0^2e^2V\frac{d}{dt}\int_{0}^{\infty}
d\tau \left\langle[\hat{\cal O}(\tau),\hat{\cal O}(0)]\right\rangle_0=0.
\]
The remaining contribution $I = \bigl\langle\hat I_0\bigr\rangle$ 
corresponds to tunneling current between two bulk reservoirs 
containing $1$- and $2$-particles. Its evaluation follows the standard
procedure~\cite{Mahan} and yields~\cite{Erice}
\beq
\frac{dI}{dV} = G_0 \frac{1}{2} \sum_{s}
\left[- \pi\nu \im T_s (eV)\right] 
\label{6.24}
\eeq
for the differential conductance across the dot at zero temperature. 
Here $G_0$ coincides with the small $\theta_0$-limit of \Eq{6.9}. 
Using now Eqs.~\eqref{4}, \eqref{5}, and~\eqref{6.20}, we obtain 
\beq
\frac{1}{G_0} \frac{dI}{dV}
=  \left\{\begin{array}{lr}
 1- \displaystyle\frac{3}{2}\left(\frac{eV}{T_K}\right)^2, 
 & |eV|\ll T_K \\ \\
\displaystyle\frac{3\pi^2/16}{\ln^2(|eV|/T_K)}\,, 
 & |eV| \gg T_K 
\end{array}
\right.
\label{6.25}
\eeq
Thus, a large voltage bias has qualitatively the same destructive 
effect on the Kondo physics as the temperature does. The 
result~\Eq{6.25} remains valid as long as $T\ll |eV|\ll \delta E$. 
If temperature exceeds the bias, $T\gg eV$, the differential 
conductance coincides with the linear conductance, see 
Eqs.~\eqref{6.12} and \eqref{6.21} above.

\subsection{Splitting of the Kondo peak in a magnetic field}
\label{split_peaks}

According to \Eq{6.25}, the differential conductance exhibits a 
peak at zero bias. The very appearance of such peak, however, does 
not indicate that we are dealing with the Kondo effect. Indeed, 
even in the absence of interactions, tunneling via a resonant level 
situated at the Fermi energy would also result in a zero-bias peak in $dI/dV$. 
These two situations can be distinguished by considering the evolution 
of the zero-bias peak with the magnetic field. In both cases, a sufficiently
strong field (such that the corresponding Zeeman energy $B$ appreciably 
exceeds the peaks width) splits the zero-bias peak in two smaller ones, 
located at $\pm V^*$.  In the case of conventional resonant tunneling, 
$eV^* = B/2$. However, if the split peaks are of the Kondo origin, 
then $eV^* \approx B$. This doubling of the distance between the split 
peaks is viewed by many as a hallmark of the many-body physics 
associated with the Kondo effect~\cite{kondo_exp}.
 
In order to address the splitting of the zero-bias Kondo peak we 
add to the Hamiltonian~\eqref{5.12} a term 
\beq
H_B=BS_z+\sum_{ks}\frac{Bs}{2}\psi_{1ks}^\dagger\psi_{1ks}
\label{6.50}
\eeq
describing the Zeeman effect of the magnetic field\footnote{In
general, magnetic field affects both the orbital and spin parts of
the electronic wave function. However, as long as spin of the dot
remains equal $1/2$ (the opposite case is considered in the next Section),
the orbital effect of the field is unimportant. It modifies the exchange 
amplitude $J$ and the bandwidth $\Delta$ in the Hamiltonian~\eqref{5.12}, 
which may eventually alter   the values of $G_0$ and $T_K$ in \Eq{6.25}.  
However, the functional   form of the dependence of $dI/dV$ on $V$ 
at a fixed $B$ remains intact.}.  We concentrate on the most interesting 
case of
\beq 
T_K\ll B\ll\Delta.
\label{6.51}
\eeq
The first inequality here ensures that the zero-bias peak is split, 
while the second one allows the development of the Kondo 
correlations to a certain extend. For simplicity, we also consider 
here a strongly asymmetric setup, as in Section~\ref{NONEQ}, 
which allows us to use \Eq{6.24} relating the differential conductance 
to the t-matrix. Note, however, that the results presented below remain 
qualitatively correct even when this assumption is lifted.

As long as the bias $|eV|$ is small compared to $B$, the differential 
conductance approximately coincides with the linear one, see 
\Eq{6.21c}. Similarly, at $|eV|\gg B$ the field has a negligible effect, 
and $dI/dV$ is given by the second line in \Eq{6.25}. The presence 
of the field, however, significantly affects the dependence of $dI/dV$ 
on $V$ at $|eV| \sim B\gg T$, and it is this regime that we address here.

\begin{figure}[h]
\centerline{\includegraphics[width=0.95\textwidth]{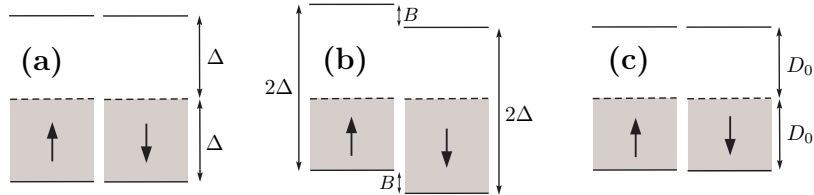}}
\vspace{2mm}
\caption{The energy bands for spin-up and spin-down
itinerant electrons. The solid lines denote the boundaries 
of the bands, the dashed lines correspond to the Fermi energy.
\newline
(a) Band structure without Zeeman splitting; (b) same with $B\neq 0$;
(c) same after the initial asymmetric cutting of the bands $(D_0 = \Delta-B)$.}
\label{band}
\end{figure}

Clearly, for $B\sim|eV|\gg T_K$ the system is in the weak coupling 
Kondo regime, so that the perturbative RG procedure described in 
Section~\ref{weak_coupling} is an adequate tool to study it. 
Let us suppose, for simplicity, that in the absence of magnetic field 
electrons are filling the lower half of a band characterized by width 
$2\Delta$ and constant density of states $\nu$, see Fig.~\ref{band}(a).  
In the presence of the field, the energy of an itinerant electron consists 
of the orbital part $\xi_k$ and the Zeeman energy $Bs/2$, see the second 
term in \Eq{6.50}. Condition $\xi_k+Bs/2 = \epsilon_F$ cuts out
different strips of orbital energies for spin-up and spin-down
electrons, see Fig.~\ref{band}(b). 
To set the stage for the RG, it is convenient to cut the initial band 
asymmetrically for spin-up and spin-down electrons. The resulting band 
structure, shown in Fig.~\ref{band}(c), is the same for both directions 
of spin. Further symmetric reductions of the band in the course of RG 
leave this band structure intact.  

After the initial cutting of the band, the effective Hamiltonian 
$H_{\rm ef\mbox{}f} + H_B$ assumes the form 
\beq
H =\sum_{ks} \xi^\pdag_k 
{\psi}^\dagger_{ks}{\psi}^\pdag_{ks}
+ J({\bf s}_\psi \cdot{\bf S}) + \eta B S^z .
\label{6.52}
\eeq
Here $k$ is measured from the Fermi momentum of electrons 
with spin $s$, and we suppressed the subscript $1$ in the operators 
$\psi_{1ks}$ etc. The Hamiltonian \eqref{6.52} is defined within a 
symmetric band 
\[
|\xi_k|< D_0 = \Delta -B\approx \Delta,
\]
see Fig.~\ref{band}(c). 
The parameter $\eta$ in \eqref{6.52} is given by
\beq
\eta = 1-\nu J/2
\label{6.53}
\eeq
with $J$ being the ``bare'' value of the exchange constant. The
correction $-\nu J/2$ in Eq.~(\ref{6.53}) is nothing else but the 
Knight shift~\cite{Knight}. In a finite magnetic field, there are 
more spin-down electrons than spin-up ones, see Fig.~\ref{band}(b). 
This mismatch results in a non-zero ground state expectation value 
$\langle s_1^z\rangle = -\nu B/2$ of the operator $s_1^z$ in \Eq{5.12}.  
Due to the exchange interaction, this induces an additional effective 
magnetic field $-\nu JB/2$ acting on the localized spin.

The form of the effective Hamiltonian \eqref{6.52} remains 
invariant under the RG transformation. The evolution of $J$ and $\eta$ 
with the reduction of the bandwidth $D$ is governed by the equations
\beqa
&&\frac{dJ_D}{d\zeta} =\nu J_D^2,
\label{6.54} \\
&&\frac{d\ln\eta_D}{d\zeta}=-\frac{1}{2}(\nu J_D)^2,
\label{6.55} 
\eeqa
where $\zeta = \ln\left({\Delta}/{D}\right)$.
Equation~\eqref{6.54} was derived in Section~\ref{weak_coupling} 
above; the derivation of \Eq{6.55} proceeds in a similar way.
It is clear from the solution of \Eq{6.54}, see \Eq{8a}, that $\nu J_D$
remains small throughout the weak coupling regime $D\gg T_K$. 
It then follows from Eqs.~\eqref{6.53} and \eqref{6.55} that 
$\eta_D$ is of the order of $1$ in this range of $D$. The 
renormalization described by Eqs.~\eqref{6.54} and \eqref{6.55} 
must be terminated at $D\sim 2B\gg T_K$, so that $\eta_B B$ 
remains smaller than $D$. 
At this point 
\beq
\nu J_B = \frac{1}{\ln (B/T_K)}\,,
\label{6.56}
\eeq
see \Eq{8a}. Solution of Eq.~(\ref{6.55}) in the scaling 
limit\footnote{
In this limit the Knight shift in \Eq{6.53} can be neglected.
}, defined as
\[
\frac{\ln(B/T_K)}{\ln(\Delta/T_K)} \ll 1,
\] 
takes the familiar form~\cite{Kondo_reviews} 
\beq
\eta_B =1-\frac{1}{2\ln(B/T_K)}\,.
\label{6.57}
\eeq

Now we are ready to evaluate the differential conductance.
The Hamiltonian Eq.~(\ref{6.52}) with $J$ and $\eta$ given by 
Eqs.~\eqref{6.56} and \eqref{6.57} is defined in a sufficiently wide
band $D\sim 2B$ to allow for excitations with energy of the order
of Zeeman splitting. Since the current operator remains invariant in 
the course of renormalization, the differential conductance can still 
be calculated from \Eq{6.24}. In the lowest order in $\nu J_B$ 
(Born approximation) we find
\beq 
\frac{1}{G_0}\frac{dI}{dV} =
\frac{\pi^2/16}{\ln^2(B/T_K)}\Bigl[1+2\theta(|eV|-\eta_B B)\Bigr].
\label{6.58}
\eeq 
The main new feature of \Eq{6.58} compared to \eqref{6.25} 
is a threshold behavior at $|eV| = \eta_B B$. The origin of this behavior 
is apparent: for $|eV|>\eta_B B$ an inelastic (spin-flip) scattering
channel opens up, causing a step in $dI/dV$. 

In the absence of Zeeman splitting, going beyond the Born
approximation in the renormalized exchange amplitude \Eq{8a} 
would be meaningless, as this would exceed the accuracy of 
the logarithmic RG. This largely remains true here as well, except 
when the bias is very close to the inelastic scattering threshold,
\[
\Bigl||eV| - \eta_B B\Bigr|\ll B.
\]
The existence of the threshold makes the t-matrix singular 
at $\omega=\eta_B B$; such logarithmically divergent contribution 
appears in the third order in $J_B$~\cite{Appelbaum}. Retaining 
of this contribution does not violate the accuracy of our 
approximations, and we find
\beqa 
\frac{1}{G_0}\frac{dI}{dV} &=&
\frac{\pi^2/16}{\ln^2(B/T_K)}\Bigl[1+2\theta(|eV|-\eta_B B)\Bigr]
\nonumber\\
&&~~~~~~~~~~~~~
\times\left(1+\frac{1}{\ln(B/T_K)}\ln\frac{B}{\bigl||eV|-\eta_B B\bigr|}\right).
\label{6.59}
\eeqa 
We see now that  the zero-bias peak in the differential
conductance is indeed split in two by the applied magnetic field. 
The split peaks are located at $\pm V^*$ with
\beq
V^* = \eta_B B = \left[1-\frac{1}{2\ln(B/T_K)}\right]B\,,
\label{6.60}
\eeq
see Eq.~(\ref{6.57}).

Because of the logarithmic divergence in \Eq{6.59} this result needs
a refinement. A finite temperature $T\gg T_K$ would of course cut
the divergence~\cite{Appelbaum}; this amounts to the replacement of 
$\bigl||eV|-\eta_B B\bigr|$ 
under the logarithm in \Eq{6.59} by 
${\rm max}\Bigl\{\bigl||eV|-\eta_B B\bigr|, T\Bigr\}$. 
This is fully similar to the cut-off of the zero-bias Kondo singularity 
in the absence of the field. 

The $T=0$ case, however, is different, and turns out to be much 
simpler than the $B=0$ Kondo anomaly at zero bias.  The logarithmic 
singularity in \Eq{6.59} is brought about by the second-order in $J_B$ 
contribution to the scattering amplitude. This contribution involves a 
transition to the excited $(S^z=+1/2)$ state of the localized spin. 
Unlike the $B=0$ case, now such a state has a finite lifetime: 
the spin of the dot may flip, exciting (a triplet) electron-hole pair 
of energy $\eta_B B$ in the band. Such relaxation mechanism was 
first considered by Korringa~\cite{Korringa}. In our case the 
corresponding relaxation rate is easily found with the help of the 
Fermi Golden Rule,
\beq
\gamma_K 
= \frac{\pi}{2} \frac{\eta_B B}{\ln^2(B/T_K)}.
\label{6.61}
\eeq
Due to the finite Korringa relaxation rate the state responsible for 
the logarithmic singularity is broadened by $\gamma_K$, which 
cuts off the divergence~\cite{LW} in \Eq{6.59}. It also smears 
the step-like dependence on $V$, which we found in the Born 
approximation, see \Eq{6.58}. As a result, the split peaks in the 
$V-$dependence of $dI/dV$ are broad and low. The width of 
the peaks is
\beq
\delta V \sim \gamma_K\sim \frac{B}{\ln^2(B/T_K)}\,,
\label{6.62}
\eeq
and their height is given by
\begin{equation}
\frac{G(V^*)-G(0)}{G(0)}
\sim\frac{\ln\left[\ln(B/T_K)\right]}{\ln(B/T_K)}
\ll 1.
\label{6.63}
\end{equation}
Here we used the short-hand notation $dI/dV = G(V)$.  When evaluating
$G(V^*)$, we replaced $\bigl||eV|-\eta_B B\bigr|$ by $\gamma_K$ in
the argument of the logarithm in \Eq{6.59}.  The right-hand side of
\Eq{6.63} is small, and the higher-order in $\nu J_B$ corrections to
it are negligible.

The problem of the field-induced splitting of the Kondo peak was revisited 
many times since the works of Appelbaum~\cite{Appelbaum}, 
see e.g.~\cite{MWL,Moore,Paaske}. 

\subsection{Kondo effect in quantum dots with large spin}
\label{large_S}

If the dot's spin exceeds $1/2$, see Refs.~\cite{weis,Leo,Kogan},
then, as discussed in Section~\ref{H_K} above, both exchange constants
$J_\gamma$ in the effective Hamiltonian~\eqref{5.2} are finite and
positive. This turns out to have a dramatic effect on the dependence
of the conductance in the Kondo regime on temperature $T$ and on
Zeeman energy $B$. Unlike the case of $S=1/2$ on the dot, see
Fig.~\ref{overall}, now the dependences on $T$ and $B$ are
\textit{non-monotonic}: initial increase of $G$ follows by a drop when
the temperature is lowered~\cite{real,ISS} at $B=0$; the variation of
$G$ with $B$ at $T=0$ is similarly non-monotonic.
 
The origin of this peculiar behavior is easier to understand by 
considering the $B$-dependence of the zero-temperature 
conductance~\cite{real}. We assume that the magnetic field 
$H_{\parallel}$ is applied \textit{in the plane} of the dot. Such
field leads to the Zeeman splitting $B$ of the spin states of the 
dot, see \Eq{6.17}, but barely affects the orbital motion of electrons.

At any finite $B$ the ground state of the system is not degenerate. 
Therefore, the linear conductance at $T = 0$ can be calculated from 
the Landauer formula
\beq
G = \frac{e^2}{h} \sum_s \left| \mathbb{S}_{s;RL}^2 \right|,
\label{7.1}
\eeq
which relates $G$ to the amplitude of scattering $\mathbb{S}_{s;RL}$ 
of an electron with spin $s$ from lead $L$ to lead $R$. The 
amplitudes $\mathbb{S}_{s;\alpha\alpha'}$ form $2\times 2\,$ 
scattering matrix $\hat{\mathbb S}_s$. In the basis of  ``channels'', 
see \Eq{5.2}, this matrix is obviously diagonal, and its eigenvalues 
$\exp\negthinspace\left(2i\delta_{\gamma s}\right)$ are related to 
the scattering phase shifts $\delta_{\gamma s}$. The scattering 
matrix in the original $(R-L)$ basis is obtained from
\[
\hat{\mathbb{S}}_s = \hat U^\dagger 
{\rm diag} \left\{e^{2i\delta_{\gamma s}}\right\} 
\hat U,
\]
where $\hat U$ is a matrix of a rotation by an angle $\theta_0$, see \Eq{6.6}.
The Landauer formula~\eqref{7.1} then yields
\beq
G = G_0 \frac{1}{2} \sum_s \sin^2\left(\delta_{1s} - \delta_{2s}\right),
\quad 
G_0 = \frac{2e^2}{h}\sin^2(2\theta_0) ,
\label{7.2}
\eeq
which generalizes the single-channel expression~\eqref{6.14}. 

Equation~\eqref{7.2} can be further simplified for a particle-hole symmetric 
model \eqref{5.2}. Indeed, in this case the phase shifts satisfy 
$\delta_{\gamma \uparrow} + \delta_{\gamma\downarrow} = 0$,
cf.~\Eq{6.16}, which suggests a representation
\beq
\delta_{\gamma s} = s\delta_\gamma .
\label{7.3a}
\eeq
Substitution into~\Eq{7.2} then results in
\beq
G = G_0 \sin^2\left(\delta_1 - \delta_2\right) .
\label{7.3}
\eeq

\begin{figure}[h]
\centerline{\includegraphics[width=0.7\textwidth]{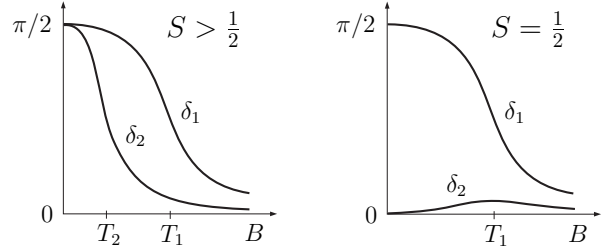}}
\vspace{2mm}
\caption{Dependence of the scattering phase shifts at the Fermi level 
on the magnetic field for $S>1/2$ (left panel) and $S=1/2$ (right panel).
}
\label{phase_shifts}
\end{figure}

\begin{figure}[h]
\centerline{\includegraphics[width=0.35\textwidth]{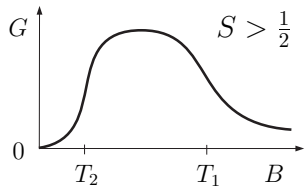}}
\vspace{2mm}
\caption{Sketch of the magnetic field dependence of the Kondo 
contribution to the linear conductance at zero temperature. The 
conductance as function of temperature exhibits a similar non-monotonic 
dependence.}
\label{G_large_S}
\end{figure}

If spin on the dot $S$ exceeds $1/2$, then both channels of itinerant
electrons participate in the screening of the dot's spin~\cite{NB}.
Accordingly, in the limit $B\to 0$ both phase shifts $\delta_\gamma$ 
approach the unitary limit value $\pi/2$, see Fig.~\ref{phase_shifts}.  
However, the increase of the phase shifts with lowering the field is 
characterized by two different energy scales. These scales, the Kondo 
temperatures $T_1$ and $T_2$, are related to the corresponding 
exchange constants in the effective Hamiltonian~\eqref{5.2},
\[
\ln\left(\frac{\Delta}{\,T_\gamma}\right) \sim \frac{1}{\nu J_\gamma} ,
\]
so that $T_1>T_2$ for $J_1 > J_2$. It is then obvious from \Eq{7.3} 
that the conductance across the dot is small both at weak $(B\ll T_2)$ 
and strong $(B\gg T_1)$ fields, but may become large $(\sim G_0)$ 
at intermediate fields $T_2\ll B \ll T_1$, see Fig.~\ref{G_large_S}. 
In other words, the dependence of zero-temperature conductance on 
the magnetic field is non-monotonic.

This non-monotonic dependence is in sharp contrast with the
monotonic increase of the conductance with lowering the field 
when $S = 1/2$. Indeed, in the latter case it is the channel whose 
coupling to the dot is the strongest that screens the dot's spin, 
while the remaining channel decouples at low energies~\cite{NB}, 
see Fig.~\ref{phase_shifts}. 

The dependence of the conductance on temperature $G(T)$ is very
similar\footnote{Note, however, that $ \langle
  \psi^\dagger_1(t)\psi_2^\pdag (t)
  \psi_2^\dagger(0)\psi_1^\pdag(0)\rangle \neq \langle
  \psi^\dagger_1(t)\psi_1^\pdag(0)\rangle \langle \psi_2^\pdag(t)
  \psi_2^\dagger(0)\rangle $ at finite $T$.  Therefore,
  unlike~\Eq{6.14}, there is no simple generalization of \Eq{6.8} to
  the two-channel case.}  to $G(B)$.  For example, for $S = 1$ one
obtains~\cite{real}
\beq
G/G_0 = \left\{
\begin{array}{lr}
\displaystyle
 (\pi T)^2 \left(\frac{1}{T_1} - \frac{1}{T_2}\right)^2, & T\ll T_2 
 \\ 
\displaystyle
\frac{\pi ^2}{2} 
\left[\frac{1}{\ln(T/T_1)} - \frac{1}{\ln(T/T_2)}\right]^2,  & T\gg T_1 
\end{array}
\right.
\label{7.4}
\eeq
The conductance reaches its maximun $G_{\rm max}$ 
at $T\sim\sqrt{T_1 T_2}$. The value of $G_{\rm max}$ can 
be found analytically for $T_1 \gg T_2$. For $S = 1$ 
the result reads~\cite{real}
\beq
G_{\rm max} = G_0 \left[1- \frac{3\pi^2}{\ln^2(T_1/T_2)}\right] .
\label{7.5}
\eeq

Consider now a Coulomb blockade valley with $N = {\rm even}$ 
electrons and spin $S=1$ on the dot. In a typical situation, the 
dot's spin in two neighboring valleys (with $N\pm 1$ electrons) is $1/2$. 
Under the conditions of applicability of the approximation~\Eq{5.5}, 
there is a single non-zero exchange constant $J_{N\pm 1}$ for each 
of these valleys. If the Kondo temperatures $T_K$ are the same 
for both valleys with $S=1/2$, then $J_{N+1}=J_{N-1}=J_{\rm odd}$. 
Each of the two singly-occupied energy levels in the valley with $S=1$ 
is also singly-occupied in one of the two neighboring valleys. 
It then follows from Eqs.~\eqref{5.4}-\eqref{5.6} that the 
exchange constants $J_{1,2}$ for $S=1$ satisfy
\[
J_1 + J_2 = \frac{1}{2} \left(J_{N+1}+J_{N-1}\right) = J_{\rm odd}.
\]
Since both $J_1$ and $J_2$ are positive, this immediately implies that
$J_{1,2} < J_{\rm odd}$. 
Accordingly, both Kondo temperatures $T_{1,2}$ are expected 
to be smaller than $T_K$ in the nearby valleys with $S = 1/2$.

This consideration, however, is not applicable when the dot is 
tuned to the vicinity of the singlet-triplet transition in its ground 
state~\cite{Sasaki,induced_review,Leo,Kogan}, i.e. when the 
energy gap $\Delta$ between the triplet ground state and the singlet 
excited state of an isolated dot is small compared to the mean level 
spacing $\delta E$. In this case the exchange constants in the 
effective Hamiltonian~\eqref{5.2} should account for additional 
renormalization that the system's parameters undergo~\cite{ST} when the 
high-energy cutoff (the bandwidth of the effective Hamiltonian) $D$ 
is reduced from $D\sim\delta E$ down to $D\sim \Delta\ll\delta E$, 
see also~\cite{PG}. The renormalization enhances the exchange 
constants $J_{1,2}$. If the ratio $\Delta/\delta E$ is sufficiently small, 
then the Kondo temperatures $T_{1,2}$ for $S=1$ 
may become of the same order~\cite{weis,Kogan}, or even 
significantly exceed~\cite{Sasaki,induced_review,Leo} the 
corresponding scale $T_K$ for $S=1/2$.

In GaAs-based lateral quantum dot systems the value of $\Delta$ can be
controlled by a magnetic field $H_\perp$ applied \textit{perpendicular
to the plane} of the dot~\cite{Leo}. Because of the smallness of the
effective mass $m^\ast$, even a weak field $H_\perp$ has a strong
orbital effect. At the same time, smallness of the quasiparticle g-factor 
in GaAs ensures that the corresponding Zeeman splitting remains 
small~\cite{induced_review}. Theory of the Kondo effect in
lateral quantum dots in the vicinity of the singlet-triplet transition
was developed in~\cite{ST_lateral}, see also~\cite{Zeeman}.

\section{Concluding remarks}

In the simplest form of the Kondo effect considered in these notes, 
a quantum dot behaves essentially as an artificial ``magnetic impurity'' 
with spin $S$, coupled via exchange interaction to two conducting 
leads. The details of the temperature dependence $G(T)$ of the 
linear conductance across the dot depend on the dot's spin~$S$. 
In the most common case of $S=1/2$ the conductance in the Kondo 
regime monotonically increases with the decrease of temperature, 
potentially up to the quantum limit $2e^2/h$. Qualitatively (although 
not quantitatively), this increase can be understood from the Anderson 
impurity model in which the dot is described by a single energy level. 
On the contrary, when spin on the dot exceeds $1/2$, the evolution 
of the conductance proceeds in two stages: the conductance first raises, 
and then drops again when the temperature is lowered.

In a typical experiment~\cite{kondo_exp}, one measures the 
dependence of the differential conductance on temperature $T$, 
Zeeman energy $B$, and dc voltage bias $V$. When one of these 
parameters is much larger than the other two, and is also large compared 
to the Kondo temperature $T_K$, the differential conductance exhibits 
a logarithmic dependence
\beq
\frac{1}{G_0} \frac{dI}{dV}
\propto \left[\ln \frac{\mathrm{max} \left\{T,B,eV\right\}}{T_K} \right]^{-2},
\label{8.1}
\eeq
characteristic for the weak coupling regime of the Kondo system,
see Section~\ref{weak_coupling}. 
Consider now a zero-temperature transport through a quantum dot 
with $S=1/2$ in the presence of a strong field $B\gg T_K$. In 
accordance with~\Eq{8.1}, the differential conductance is small 
compared to $G_0$  both for $eV\ll B$ and for $eV\gg B$. However, 
the calculation in the third order of perturbation theory in the exchange 
constant yields a contribution that diverges logarithmically at 
$eV \approx \pm B$~\cite{Appelbaum}. The divergence is reminiscent 
of the Kondo zero-bias anomaly. 
However, the full development of resonance is inhibited by a finite 
lifetime of the excited spin state of the dot~\cite{LW}. As a 
result, the peak in the differential conductance at $eV = \pm \mu B$ 
is broader 
and lower then the corresponding peak at zero bias in 
the absence of the field, see Section~\ref{split_peaks}. 

One encounters similar effects in studies of the influence of a weak 
ac signal of frequency $\Omega\gtrsim T_K$ applied to the gate 
electrode~\cite{Elzerman} on transport across the dot. In close analogy 
with the usual photon-assisted tunneling~\cite{TG}, such perturbation is 
expected to result in the formation of satellites~\cite{HS} at 
$eV = n\hbar\Omega$ (here $n$ is an integer) to the zero-bias peak 
in the differential conductance. Again, the resonances
responsible for the formation of the satellite peaks are limited by the finite 
lifetime effects~\cite{KNG}.  

The spin degeneracy is not the only possible source of the Kondo 
effect in quantum dots. Consider, for example, a large dot connected 
by a single-mode junction to a conducting lead and tuned to the vicinity 
of the Coulomb blockade peak~\cite{KM}. If one neglects the finite 
level spacing in the dot, then the two almost degenerate charge state 
of the dot can be labeled by a pseudospin, while real spin plays the 
part of the channel index~\cite{KM,Matveev}. This setup turns out 
to be a robust realization~\cite{KM,Matveev} of the symmetric (i.e. 
having equal exchange constants) two-channel $S=1/2$ Kondo 
model~\cite{NB}. The model results in a peculiar temperature dependence 
of the observable quantities, which at low temperatures follow power 
laws with manifestly non-Fermi-liquid fractional values of the 
exponents~\cite{CZ}. 

It should be emphasized that in the usual geometry consisting of two 
leads attached to a Coulomb-blockaded quantum dot with $S=1/2$, 
only the conventional Fermi-liquid behavior can be observed at low 
temperatures. Indeed, in this case the two exchange constants in the 
effective exchange Hamiltonian~\eqref{5.2} are vastly different, and 
their ratio is not tunable by conventional means, see the discussion in 
Section~\ref{H_K} above. A way around this difficulty was proposed 
in~\cite{OGG}. The key idea is to replace one of the leads in 
the standard configuration by a very large quantum dot, characterized 
by a level spacing $\delta E'$ and a charging energy $E_C^\prime$. 
At $T\gg \delta E'$, particle-hole excitations within this dot are allowed, 
and electrons in it participate in the screening of the smaller dot's spin. 
At the same time, as long as $T\ll E_C^\prime$, the number of electrons 
in the large dot is fixed. Therefore, the large dot provides for a separate 
screening channel which does not mix with that supplied by the remaining 
lead. In this system, the two exchange constants are controlled by the 
conductances of the dot-lead and dot-dot junctions. A strategy for tuning 
the device parameters to the critical point characterized by the two-channel 
Kondo physics is discussed in~\cite{2CK}. 

Finally, we should mention that the description based on the universal 
Hamiltonian~\eqref{2.12} is not applicable to large quantum dots subjected 
to a {\it quantizing} magnetic field $H_\perp$~\cite{checkers_1,checkers_2}. 
Such field changes drastically the way the screening occurs in a confined 
droplet of a two-dimensional electron gas~\cite{edge}. The droplet is 
divided into alternating domains containing compressible and incompressible 
electron liquids. In the metal-like compressible regions, the screening is almost 
perfect. On the contrary, the incompressible regions behave very much 
like insulators. In the case of lateral quantum dots, a large compressible 
domain is formed near the center of the dot. The domain is surrounded 
by a narrow incompressible region separating it from another compressible 
ring-shaped domain formed along the edges of the dot~\cite{rings}. This 
system can be viewed as two concentric capacitively coupled quantum 
``dots" - the core dot and the edge dot~\cite{checkers_1,rings}. When 
the leads are attached to the edge dot, the measured conductance is sensitive 
to its spin state: when the number of electrons in the edge dot is odd, 
the conductance becomes large due to the Kondo effect~\cite{checkers_1}. 
Changing the field causes redistribution of electrons between the core and 
the edge, resulting in a striking checkerboard-like pattern of high- and 
low-conductance regions~\cite{checkers_1,checkers_2}. This behavior 
persists as long as the Zeeman energy remains small compared to the 
Kondo temperature. Note that compressible regions are also formed 
around an {\it antidot} -- a potential hill in a two-dimensional electron 
gas in the quantum Hall regime~\cite{GS}. Both Coulomb blockade oscillations 
and Kondo-like behavior were observed in these systems~\cite{antidot}.

Kondo effect arises whenever a coupling to a Fermi gas induces
transitions within otherwise degenerate ground state multiplet of an
interacting system. Both coupling to a Fermi gas and interactions are
naturally present in a nanoscale transport experiment. At the same
time, many nanostructures can be easily tuned to the vicinity of a
degeneracy point. This is why the Kondo effect in its various forms 
often influences the low temperature transport in meso- and nanoscale
systems. 

In these notes we reviewed the theory of the Kondo effect in transport 
through quantum dots. A Coulomb-blockaded quantum dot behaves 
in many aspects as an artificial ``magnetic impurity'' coupled via exchange 
interaction to two conducting leads. Kondo effect in transport through 
such ``impurity'' manifests itself in the lifting of the Coulomb blockade 
at low temperatures, and, therefore, can be unambiguously identified. 
Quantum dot systems not only offer a direct access to transport properties 
of an artificial impurity, but also provide one with a broad arsenal of tools 
to tweak the impurity properties, unmatched in conventional systems. The 
characteristic energy scale for the intra-dot excitations is much smaller 
than the corresponding scale for natural magnetic impurities. This allows 
one to induce degeneracies in the ground state of a dot which are more 
exotic than just the spin degeneracy. This is only one out of many possible 
extensions of the simple model discussed in these notes.

\section{Acknowledgements}
This project was supported by NSF grants DMR02-37296 and EIA02-10736, 
and by the Nanoscience/Nanoengineering Research Program of Georgia Tech.



\begin{thebibliography}{99}

\bibitem{blockade}
L.P. Kouwenhoven et al., 
in: {\it Mesoscopic Electron Transport}, 
eds. L.L. Sohn et al., 
(Kluwer, Dordrecht, 1997), p.~105;
M.A. Kastner, \RMP {\bf 64} (1992) 849;
U. Meirav and E.B. Foxman, 
Semicond. Sci. Technol. {\bf 11} (1996) 255;
L.P. Kouwenhoven and C.M. Marcus,
Phys. World {\bf 11} (1998) 35.

\bibitem{Devoret}
P. Joyez et al.,  \PRL {\bf 79} (1997) 1349;
M. Devoret and C. Glattli, Phys. World {\bf 11} (1998) 29.

\bibitem{Altsuler_Aronov} 
B.L. Altshuler and A.G. Aronov, in:
{\it Electron-Electron Interactions in Disordered Systems}, eds. A.L. Efros and M. Pollak
(North-Holland, Amsterdam, 1985), p.~ 1.

\bibitem{kondo_exp}  
D. Goldhaber-Gordon et al.,   
Nature {\bf 391} (1998) 156; 
S.M. Cronenwett, T.H. Oosterkamp and L.P. Kouwenhoven,
Science {\bf 281} (1998) 540; 
J. Schmid et al.,  
Physica B {\bf 256-258} (1998) 182.

\bibitem{kondo_popular}
L. Kouwenhoven and L. Glazman, Physics World {\bf 14} (2001) 33.

\bibitem{Kondo} 
J. Kondo, Prog. Theor. Phys. {\bf 32} (1964) 37.

\bibitem{FM} 
Furusaki A and K.A. Matveev
Phys. Rev. B {\bf 52} (1995) 16676.

\bibitem{AG1}
I.L. Aleiner and L.I. Glazman,
Phys. Rev. B {\bf 57} (1998) 9608.

\bibitem{Cronenwett}
S.M. Cronenwett et al., \PRL {\bf 81} (1998) 5904.

\bibitem{vertical} 
S. Tarucha et al., 
\PRL {\bf 84} (2000) 2485; 
L.P. Kouwenhoven, D.G. Austing and S. Tarucha, 
Rep. Prog. Phys. {\bf 64} (2001) 701.

\bibitem{Sasaki}
S. Sasaki et al., 
Nature {\bf 405} (2000) 764;
S. Sasaki et al., 
\PRL {\bf 93} (2004) 017205.

\bibitem{induced_review}
M. Pustilnik et al., 
Lecture Notes in Physics {\bf 579} (2001) 3 (cond-mat/0010336).

\bibitem{nanotube}
J. Nyg{\aa}rd, D.H. Cobden and P.E. Lindelof,
Nature {\bf 408} (2000) 342;
W. Liang, M. Bockrath and H. Park, \PRL {\bf 88} (2002) 126801;
B. Babi{\'c}, T. Kontos and C. Sch{\"o}nenberger, cond-mat/0407193. 

\bibitem{Park} 
J. Park et al., 
Nature {\bf 417} (2002) 722;
W. Liang  et al., 
Nature {\bf 417} (2002) 725;
L.H. Yu and D. Natelson, Nano Lett. {\bf 4} (2004) 79. 

\bibitem{mirage}
L.T. Li et al., 
\PRL {\bf 80} (1998) 2893;
V. Madhavan et al., 
Science {\bf 280} (1998) 567;
H.C. Manoharan et al., 
Nature {\bf 403} (2000) 512. 

\bibitem{RMT1} 
M.V. Berry,  Proc. R. Soc. A {\bf 400} (1985) 229;
B.L. Altshuler and B.I. Shklovskii, 
Zh. Eksp. Teor. Fiz. {\bf 91} (1986) 220
[Sov. Phys.--JETP {\bf 64} (1986) 127].

\bibitem{Beenakker_RMP}
C.W.J. Beenakker, \RMP {\bf 69} (1997) 731.

\bibitem{Alhassid_RMP}
Y. Alhassid, \RMP {\bf 72} (2000) 895.

\bibitem{Haake}
F. Haake, {\it Quantum Signatures of Chaos} (Springer-Verlag, New York, 2001).

\bibitem{Efetov}
K. Efetov, {\it Supersymmetry in Disorder and Chaos} 
(Cambridge University Press, Cambridge, 1997).

\bibitem{RMT} 
B.L. Altshuler et al.,
\PRL {\bf 78} (1997) 2803;
O. Agam et al., 
\PRL {\bf 78} (1997) 1956;
Ya.M. Blanter, \PRB {\bf 54} (1996) 12807;
Ya.M. Blanter and A.D. Mirlin, \PRB {\bf 57} (1998) 4566;
Ya.M. Blanter, A.D. Mirlin and B.A. Muzykantskii,
\PRL {\bf 78} (1997) 2449;
I.L. Aleiner and L.I. Glazman, \PRB {\bf 57} (1998) 9608.

\bibitem{KAA}
I.L. Kurland, I.L. Aleiner and B.L. Altshuler,
\PRB {\bf 62} (2000) 14886.

\bibitem{ABG}
I.L. Aleiner, P.W. Brouwer and L.I. Glazman, Phys. Rep. {\bf 358} (2002) 309.

\bibitem{wave_functions}
M.V. Berry, Journ. of Phys. A {\bf 10} (1977) 2083;
Ya.M. Blanter and A.D. Mirlin, Phys. Rev. E {\bf 55} (1997) 6514
Ya.M. Blanter, A.D. Mirlin and and B.A. Muzykantskii,
\PRB {\bf 63} (2001) 235315;
A.D. Mirlin, Phys. Rep. {\bf 326} (2000) 259.

\bibitem{Folk96}
J.A. Folk et al., \PRL {\bf 76} (1996) 1699.

\bibitem{Ziman}
J.M. Ziman, {\it Principles of the Theory of Solids}
(Cambridge University Press, Cambridge, 1972), p.~339.

\bibitem{Delft-Ralph} 
J. von Delft and D.C. Ralph, Phys. Rep. {\bf 345} (2001) 61.
 
\bibitem{MGS} 
K.A. Matveev, L.I. Glazman and R.I. Shekhter, 
Mod. Phys. Lett. B {\bf 8} (1994) 1007.

\bibitem{spin}  
P.W. Brouwer, Y. Oreg and B.I. Halperin,
\PRB {\bf 60} (1999) R13977;
H.U. Baranger, D. Ullmo and L.I. Glazman, 
\PRB {\bf 61} (2000) R2425.

\bibitem{spin_exp}
R.M. Potok et al., 
\PRL {\bf 91} (2003) 016802; 
J.A. Folk et al.,  
Phys. Scripta {\bf T90} (2001) 26; 
S. Lindemann et al., \PRB {\bf 66} (2002) 195314.

\bibitem{KM} 
K.A. Matveev, \PRB {\bf 51} (1995) 1743.

\bibitem{KF} 
K. Flensberg, \PRB {\bf 48} (1993) 11156.

\bibitem{real}
M. Pustilnik and L.I. Glazman, \PRL {\bf 87} (2001) 216601.

\bibitem{PT} 
C.E. Porter and R.G. Thomas, Phys. Rev. {\bf 104} (1956) 483.

\bibitem{rate}
I.O. Kulik and R.I. Shekhter, Zh. Eksp. Teor. Fiz. {\bf 62} (1975) 623
[Sov. Phys.--JETP {\bf 41} (1975) 308];
L.I. Glazman and R.I. Shekhter,  
Journ. of Phys.: Condens. Matter {\bf 1} (1989) 5811.


\bibitem{rate_discrete}
L.I. Glazman and K.A. Matveev,
Pis'ma Zh. Exp. Teor. Fiz. {\bf 48}, 403 (1988)
[JETP Lett. {\bf 48}, 445 (1988)];
C.W.J. Beenakker, \PRB {\bf 44} (1991) 1646;
D.V. Averin, A.N. Korotkov and K.K. Likharev, \PRB {\bf 44}, (1991) 6199. 

\bibitem{JSA}
R.A. Jalabert, A.D. Stone and Y. Alhassid, \PRL {\bf 68} (1992) 3468.

\bibitem{PEI} 
V.N. Prigodin, K.B. Efetov and S. Iida, \PRL {\bf 71} (1993) 1230.

\bibitem{Chang96} A.M. Chang et al., 
\PRL {\bf 76} (1996) 1695.

\bibitem{magnetoresistance}
B.L. Altshuler et al., 
\PRB {\bf 22} (1980) 5142; 
S. Hikami, A.I. Larkin and Y. Nagaoka, Progr. Theor. Phys.
{\bf 63} (1980) 707. 

\bibitem{Giaever}
I. Giaever and H.R. Zeller, \PRL {\bf 20} (1968) 1504;
H.R. Zeller and I. Giaever, Phys. Rev. {\bf 181} (1969) 789.

\bibitem{AN}
D.V. Averin and Yu.V. Nazarov, \PRL {\bf 65} (1990) 2446.

\bibitem{Abrikosov} 
A.A. Abrikosov, {\it Fundamentals of the Theory of Metals}
(North-Holland, Amsterdam, 1988), p.~620.

\bibitem{AG}
I.L. Aleiner and L.I. Glazman, \PRL {\bf 77} (1996) 2057.

\bibitem{AM}
L.I. Glazman and M.E. Raikh, Pis'ma Zh. Eksp. Teor. Fiz. {\bf 47} (1988) 378
[JETP Lett. {\bf 47} (1988) 452];
T.K. Ng and P.A. Lee, \PRL {\bf 61} (1988) 1768.

\bibitem{unitary}
W.G. van der Wiel et al.,
 Science {\bf 289} (2000) 2105;
Y. Ji, M. Heiblum and H. Shtrikman, \PRL {\bf 88} (2002) 076601.

\bibitem{Erice}
L.I. Glazman and M. Pustilnik, in:
{\it New Directions in Mesoscopic Physics (Towards Nanoscience)} 
eds. R. Fazio et al., (Kluwer, Dordrecht, 2003), p.~93
(cond-mat/0302159).

\bibitem{old_reviews} 
C.B. Duke, {\it Tunneling in Solids} (Academic Press, New York, 1969);
J.M. Rowell, in: {\it Tunneling Phenomena in Solids}, 
eds. E. Burstein and S. Lundqvist (Plenum Press, New York, 1969), p.~385.

\bibitem{classics} 
A.F.G. Wyatt, \PRL {\bf 13} (1964) 401;
R.A. Logan and J.M. Rowell, \PRL {\bf 13} (1964) 404.

\bibitem{Appelbaum} 
J. Appelbaum, \PRL {\bf 17} (1966) 91;
J.A. Appelbaum, Phys. Rev. {\bf 154} (1967) 633.

\bibitem{Anderson}
P.W. Anderson, \PRL {\bf 17} (1966) 95.

\bibitem{Fiete} 
G.A. Fiete et al., 
\PRB {\bf 66} (2002) 024431.

\bibitem{JPCM}
M. Pustilnik and L.I. Glazman,
Journ. of Physics: Condens. Matter {\bf 16} (2004) R513.

\bibitem{SW}
J.R. Schrieffer and P.A. Wolff, Phys. Rev. {\bf 149} (1966) 491.

\bibitem{SI}
P.G. Silvestrov and Y. Imry, \PRL {\bf 85} (2000) 2565.

\bibitem{NB} 
P. Nozi\`{e}res and A. Blandin, J. Physique {\bf 41} (1980) 193.

\bibitem{Kondo_reviews}
P. Coleman, in: 
{\it Lectures on the Physics of Highly Correlated Electron Systems VI}, 
ed. F. Mancini (American Institute of Physics, New York, 2002), p.~79
(cond-mat/0206003);
A.S. Hewson, {\it The Kondo Problem to Heavy Fermions} 
(Cambridge University Press, Cambridge, 1997).

\bibitem{PWA_book} 
P.W. Anderson, {\it Basic Notions of Condensed Matter Physics} 
(Addison-Wesley, Reading, 1997).

\bibitem{Wilson}
K.G. Wilson, \RMP {\bf 47} (1975) 773.

\bibitem{GHL} 
L.I. Glazman, F.W.J. Hekking and A.I. Larkin, 
\PRL {\bf 83} (1999) 1830.

\bibitem{bethe} 
A.M. Tsvelick and P.B. Wiegmann, Adv. Phys. {\bf 32} (1983) 453;
N. Andrei, K. Furuya and J.H. Lowenstein, \RMP {\bf 55} ( 1983) 331.

\bibitem{AL} 
I. Affleck and A.W.W. Ludwig, \PRB {\bf 48} (1993) 7297.

\bibitem{PG} 
M. Pustilnik and L.I. Glazman, \PRB {\bf 64} (2001) 045328.

\bibitem{AAA}  
A.A. Abrikosov, Physics {\bf 2} (1965) 5;
A.A. Abrikosov, Usp. Fiz. Nauk {\bf 97} (1969) 403
[Sov. Phys.--Uspekhi {\bf 12} (1969) 168];
H. Suhl, Phys. Rev. {\bf 138} (1965) A515.

\bibitem{Newton} 
R.G. Newton, {\it Scattering Theory of Waves and Particles} 
(Dover, Mineola, 2002).

\bibitem{PWA}  
P.W. Anderson, Journ. of Physics C {\bf 3} (1970) 2436.

\bibitem{AYH}
P.W. Anderson, G. Yuval and D.R. Hamann,  
\PRB {\bf 1} (1970) 4464. 

\bibitem{flow}
F. Wegner, Ann. Phys. {\bf 3} (1994) 77; 
Nucl. Phys. B {\bf 90} (2000) 141;
S.D. Glazek and K.G. Wilson,  Phys. Rev. D {\bf 49} (1994) 4214;
Phys. Rev. D {\bf 57} (1998) 3558.

\bibitem{N} 
P. Nozi\`{e}res, J. Low Temp. Phys. {\bf 17} (1974) 31;
P. Nozi{\`e}res, in \textit{Proceedings of the 14th International Conference 
on Low Temperature Physics}, eds. M. Krusius and M. Vuorio 
(North Holland, Amsterdam, 1974), Vol. 5, pp. 339-374;
P. Nozi\`{e}res, J. Physique {\bf 39} (1978) 1117.

\bibitem{Costi}
T.A. Costi, A.C. Hewson and V. Zlati\'{c},
Journ. of Physics: Condens. Matter {\bf 6} (1994) 2519.

\bibitem{Mahan} G.D. Mahan, {\it Many-Particle Physics} 
(Plenum, New York, 1990).

\bibitem{Knight} 
W.D. Knight, Phys. Rev. {\bf 76} (1949) 1259;
C. H. Townes, C. Herring, and W. D. Knight, 
Phys. Rev. {\bf 77} (1950) 736.

\bibitem{Korringa} 
J. Korringa, Physica (Amsterdam) {\bf 16} (1950) 601.

\bibitem{LW} 
D.L. Losee  and E.L. Wolf, \PRL {\bf 23} (1969) 1457.

\bibitem{MWL}
Y. Meir, N.S. Wingreen and P.A. Lee, \PRL {\bf 70} (1993) 2601.

\bibitem{Moore}
J.E. Moore and X.-G. Wen, 
Phys. Rev. Lett. {\bf 85} (2000) 1722.

\bibitem{Paaske}
J. Paaske, A. Rosch, and P. W\"olfle, 
Phys. Rev. B {\bf 69} (2004) 155330.

\bibitem{weis} 
J. Schmid et al., 
\PRL {\bf 84} (2000) 5824.

\bibitem{Leo}
W.G. van der Wiel et al., 
\PRL {\bf 88} (2002) 126803.

\bibitem{Kogan} 
A. Kogan et al., 
\PRB {\bf 67} (2003) 113309.

\bibitem{ISS}
W. Izumida, O. Sakai and Y. Shimizu, J. Phys. Soc. Jpn. {\bf 67} (1998) 2444.

\bibitem{ST} 
M. Eto and Yu.V. Nazarov, \PRL {\bf 85} (2000) 1306;
M. Pustilnik and L.I. Glazman, \PRL {\bf 85} (2000) 2993.

\bibitem{ST_lateral}
V.N. Golovach and D. Loss, Europhys. Lett. {\bf 62} (2003) 83; 
M. Pustilnik, L.I. Glazman and W. Hofstetter,
\PRB {\bf 68} (2003) 161303(R).   

\bibitem{Zeeman} 
M. Pustilnik, Y. Avishai and K. Kikoin, \PRL {\bf 84} (2000) 1756.

\bibitem{Elzerman}
J.M. Elzerman et al., 
J. Low Temp. Phys. {\bf 118} (2000) 375.

\bibitem{TG}
P.K. Tien and J.P. Gordon, Phys. Rev. {\bf 129} (1963) 647.

\bibitem{HS}
M.H. Hettler and H. Schoeller, \PRL {\bf 74} (1995) 4907. 

\bibitem{KNG} 
A. Kaminski, Yu.V. Nazarov and L.I. Glazman, 
\PRL {\bf 83} (1999) 384;
A. Kaminski, Yu.V. Nazarov and L.I. Glazman, 
\PRB {\bf 62} (2000) 8154.

\bibitem{Matveev}
K.A. Matveev, Zh. Eksp. Teor. Fiz. {\bf 99} (1991) 1598 
[Sov. Phys.--JETP {\bf 72} (1991) 892].

\bibitem{CZ}
D.L. Cox and A. Zawadowski, Adv. Phys. {\bf 47} (1998) 599.

\bibitem{OGG}
Y. Oreg and D. Goldhaber-Gordon, \PRL {\bf 90} (2003) 136602. 

\bibitem{2CK}
M. Pustilnik et al., \PRB {\bf 69} (2004) 115316.

\bibitem{checkers_1}
M. Keller et al., 
\PRB {\bf 64} (2001) 033302; 
M. Stopa et al., 
\PRL {\bf 91} (2003) 046601. 

\bibitem{checkers_2}
S.M. Maurer et al., 
\PRL {\bf 83} (1999) 1403; 
D. Sprinzak et al., 
\PRL {\bf 88} (2002) 176805; 
C. F{\"u}hner et al., 
\PRB {\bf 66} (2002) 161305(R). 

\bibitem{edge}
C.W.J. Beenakker, \PRL {\bf 64} (1990) 216;
A.M. Chang, Solid State Commun. {\bf 74} (1990) 871;
D.B. Chklovskii, B.I. Shklovskii and L.I. Glazman, 
\PRB {\bf 46} (1992) 4026.

\bibitem{rings}
P.L. McEuen et al., \PRB {\bf 45} (1992) 11419;
A.K. Evans, L.I. Glazman and B.I. Shklovskii,
\PRB {\bf 48} (1993) 11120.

\bibitem{GS}
V.J. Goldman and B. Su, {\it Science} {\bf 267} (1995) 1010.

\bibitem{antidot}
M. Kataoka et al., \PRL {\bf 83} (1999) 160;
M. Kataoka et al., 
\PRL {\bf 89} (2002) 226803.

\end{thebibliography}
\end{document}